\documentclass[11pt,a4paper]{article}
\pdfoutput=1
\usepackage{a4wide}
\usepackage{amsfonts}
\usepackage{amssymb}
\usepackage{amsmath}
\usepackage{graphicx}
\usepackage{booktabs}
\usepackage{array}
\usepackage{color}
\usepackage{ulem}
\usepackage[small,bf]{caption}
\usepackage{cite}
\usepackage[usenames,dvipsnames]{xcolor}
\usepackage{verbatim}
\usepackage{dcolumn}

\def\1{\mathbf{1}}
\def\3{\mathbf{3}}
\def\2{\mathbf{2}}

\usepackage{lineno}

\newcolumntype{d}[1]{D{.}{\cdot}{#1} }

\allowdisplaybreaks

\numberwithin{equation}{section}

\DeclareMathOperator{\diag}{Diag}
\DeclareMathOperator{\ci}{\text{i}}

\begin{document}
\begin{titlepage}

\begin{center}
{
\bf\LARGE 
Leptonic Sum Rules from\\[0.1em]
Flavour Models with Modular Symmetries
}
\\[8mm]
J.~Gehrlein$^{\, a,}$ \footnote{E-mail: \texttt{jgehrlein@bnl.gov}} and
M.~Spinrath$^{\, b,}$ \footnote{E-mail: \texttt{spinrath@phys.nthu.edu.tw}}
\\[1mm]
\end{center}
\vspace*{0.50cm}
\centerline{$^{a}$ \it High Energy Theory Group, Physics Department, Brookhaven National Laboratory,}
\centerline{\it Upton, NY 11973, USA}
\vspace*{0.2cm}
\centerline{$^{b}$ \it Department of Physics, National Tsing Hua University, Hsinchu 30013, Taiwan}
\vspace*{1.20cm}

\begin{abstract}
\noindent
Sum rules in the lepton sector provide an extremely valuable tool to classify
flavour models in terms of relations between neutrino masses and mixing parameters
testable in a plethora of experiments.
In this manuscript we identify new leptonic sum rules arising in models with modular
symmetries with residual symmetries.
These models  simultaneously present neutrino mass sum rules, involving masses and Majorana
phases, and mixing sum rules, connecting the mixing angles and the Dirac CP-violating phase.
The simultaneous appearance of both types of sum rules leads to some non-trivial interplay, for instance,
the allowed absolute neutrino mass scale exhibits a dependence on the Dirac CP-violating
phase. We derive analytical expressions for these novel sum rules and 
present their allowed parameter ranges  as well as  their predictions at upcoming neutrino experiments.
\end{abstract}

\end{titlepage}
\setcounter{footnote}{0}

\section{Introduction}

One of the big open questions of the Standard Model (SM) of particle
physics is the origin of neutrino masses and mixings. The
observation of neutrino oscillations 
demonstrated that neutrinos are massive.
However, they are predicted to be massless in the original formulation of
the SM. Since then neutrino oscillations have been well established and
the three mixing angles in the PMNS matrix have been measured to a good
accuracy (see, e.g., \cite{Esteban:2020cvm} for a global analysis of all
oscillation data), while right now we only have hints of leptonic CP
violation (CPV) \cite{Esteban:2020cvm}
and a precise
measurement of the Dirac CPV phase is up to near-future experiments.
On the other hand so far we only have an upper bound on the absolute mass
scale of neutrinos from cosmology and beta decay \cite{Tanabashi:2018oca}
which shows that neutrinos are several orders of magnitude lighter than the
other SM fermions.
 
Large theoretical effort has been devoted to understand the origin
of neutrino masses and mixings
(see \cite{Altarelli:2010gt, Ishimori:2010au, King:2013eh, King:2014nza, Feruglio:2015jfa, Petcov:2017ggy, Xing:2019vks, Feruglio:2019ktm} for recent reviews).
Particularly challenging from the theoretical point of view is  to
explain the observed pattern of neutrino mixing consisting of 
two large and one small mixing angle. 
Arguably one of the most natural explanations of this pattern  
is provided by models based on non-Abelian discrete flavour symmetries.
Testing the predictions of these models is of utmost importance 
to understand the origin of neutrino masses and mixing.
Most of the  discrete symmetry models of neutrino mixing and, more generally,
of lepton flavour, predict correlations between several observables, 
making them testable at current or upcoming experiments.
Two  different correlations between neutrino observables exist in models 
with discrete symmetries: neutrino mass sum rules
(for early papers see \cite{Altarelli:2008bg, Hirsch:2008rp, Bazzocchi:2009da, Altarelli:2009kr,Chen:2009um}, for reviews see
\cite{Barry:2010yk, Dorame:2011eb, King:2013psa, Agostini:2015dna, Gehrlein:2015ena, Gehrlein:2016wlc}),
which, 
in the case of three neutrino mixing, involve the three 
light neutrino masses and the two Majorana CPV phases 
\cite{Bilenky:1980cx} of the PMNS neutrino mixing matrix,
and leptonic mixing sum rules which relate the leptonic mixing angles and 
the Dirac CPV phase, for some early papers with such an explicit
connection, see, e.g., \cite{King:2005bj, Masina:2005hf} and for some more recent reviews
and systematic studies see, e.g., \cite{King:2013eh, King:2014nza, Hanlon:2013ska, Ge:2011qn, Ge:2011ih, Marzocca:2013cr, Petcov:2014laa, Girardi:2015vha,Girardi:2015rwa}.
Neutrino mass sum rules can be tested in experiments 
measuring the absolute neutrino mass scale 
or experiments that provide  information on the sum of  the neutrino masses.
A number of studies have shown, in particular, 
that checking the validity of the leptonic mixing sum rules using the 
currently available and prospective data on the neutrino mixing angles 
and the Dirac CPV phase is an extremely powerful method of 
discriminating between different discrete symmetry models and, more 
generally, of testing the non-Abelian discrete symmetry approach to  the
neutrino mixing problem 
(see, e.g., \cite{Agarwalla:2017wct, Petcov:2018snn, Blennow:2020snb,Blennow:2020ncm}).

Recently, a generalisation  of  the  discrete  symmetry  approach  
has  been  proposed  in \cite{Feruglio:2017spp}. In this approach modular  
invariance  plays  the role  of  the flavour  symmetry  and  couplings  
of  the theory  are  modular  forms of a  certain  level $N$.  
In the simplest realisation of this idea, the vacuum expectation value of  
a complex  field (modulus)  is  the  only  source  of  flavour  symmetry  
breaking, which leads to a reduction of free parameters in the respective 
models compared to the models based on the 
standard discrete symmetry approach where several copies of 
scalars (flavons) need to be introduced to break the flavour symmetry.
This reduction of the parameters in models with modular symmetry 
leads to a new appealing  feature,
namely, the existence of new sum rules since the neutrino  masses,  
neutrino  mixing  and the CPV 
phases  are  simultaneously determined by the modular symmetry   
typically  in  terms  of  a  limited  number  
of  constant parameters. The simultaneous presence of neutrino mass and mixing sum rules 
makes these models highly predictive and 
can be used as a target for upcoming neutrino experiments.
 
In the following we will study models based
on modular symmetries available in the literature which lead to new mass and mixing 
sum rules.
We will  focus on models which lead to the
maximal number of sum rules, i.e., models with residual symmetries. 
In particular we will concentrate on sum rules which have been
previously overlooked in these models.  
We will discuss
their predictions for the bounds on the lightest neutrino mass, the observable
in neutrinoless double beta decay, the kinematic neutrino mass as
well as the mixing parameters. Our results
can  provide a  link between model building, phenomenology, and experiments as they
allow to study which models could be distinguished by the experiment and, in the
case of an observation, the measurement can be directly linked to certain
flavour models.
 
This manuscript is organized as follows: in sec.~\ref{sec:msr} we introduce
the parametrization of mass sum rules and give general insights on the predictions
of mass sum rules, in sec.~\ref{sec:exp} we provide an explicit, detailed example for sum rules
in a model based on the modular symmetry $A_4$, sec.~\ref{sec:models} is dedicated to a
collection of models with residual symmetries present in the literature which
feature sum rules whose predictions we show in sec.~\ref{sec:results} and we
summarize and conclude in sec.~\ref{sec:sc}.
  
\section{Mass sum rules}
\label{sec:msr}

Before we look into actual models we want to discuss in some detail
mass sum rules and how we can derive certain phenomenological predictions from them.
For mixing sum rules, such a derivation is more straightforward and we will not
discuss any general statements about them here.

Mass sum rules relate the three light neutrino masses and two Majorana phases
to each other.
The existence of mass sum rules is not related to  any symmetry nor related
to a particular
mass mechanism \cite{Gehrlein:2017ryu}, it is merely the result of having
less parameters
than observables.

Using the complex mass eigenvalues $m_i \exp(- \ci \phi_i)$
neutrino mass sum rules can be generally parametrised as
\begin{equation}
\label{eq:Parameterisation}
 \begin{split}
  s(m_1,m_2,m_3,\phi_1,\phi_2, \theta_{12}, \theta_{13}, \theta_{23}, \delta, d ) & \equiv \\
   f_1(\theta_{12}, \theta_{13}, \theta_{23}, \delta)  (m_1 \, \text{e}^{-\ci \, \phi_1})^d 
 &+ f_2(\theta_{12}, \theta_{13}, \theta_{23}, \delta)  (m_2 \, \text{e}^{-\ci \, \phi_2})^d + m_3^d \stackrel{!}{=} 0 \;,
 \end{split} 
\end{equation}
where $\phi_1,~\phi_2$ are the Majorana phases  and $f_1$, $f_2$ are model
dependent, complex coefficients.
Here we resemble for easier comparison the conventions
used in \cite{Gehrlein:2015ena, Gehrlein:2016wlc}.
For  the original mass sum rules there was no explicit dependence of $f_1$ and $f_2$
on the mixing angles or the Dirac CPV phase. As we will see in the following this will not
be the case anymore for sum rules in models with modular symmetries. Here the coefficients
are functions of the mixing parameters leading to a strikingly different phenomenology of
mass sum rules in models with and without modular symmetries.
Starting from the parametrisation  in eq.~\eqref{eq:Parameterisation} we will discuss in the 
following  how one can derive expressions for observables which are affected by the
existence of a mass sum rule.

\subsection{Observables}

We want to discuss here some general formulas which will be useful
to derive constraints on observables in models which have a mass sum rule.
As  mass sum rules involve the Majorana phases the ideal observable to test
them is the observable in neutrinoless double beta decay $|m_{ee}|$. However,
mass sum rules also provide a constraint on the absolute neutrino mass scale.

\subsubsection{Neutrino mass scale}
\label{sec:vMassScale}

We begin with a formula for the lower and
upper bound on neutrino masses assuming the presence of a mass sum rule.
From experiments we have
information on two neutrino mass squared differences hence 
it is possible to rewrite two of the masses as
\begin{align}
\label{eq:massspl}
m_2 = \sqrt{m_1^2 + \Delta m_{21}^2},~m_3 = \sqrt{m_1^2 + \Delta m_{31}^2} \;,
\end{align}
which applies to both mass orderings. Normal mass ordering (NO) means: $m_1<m_2<m_3$ and 
inverted mass ordering (IO): $m_3<m_1<m_2$.

Looking at the mass sum rule from a geometrical point of view as a
triangle in the complex plane, cf.\ \cite{Dorame:2011eb, Barry:2010yk},
it is clear that the most extremal masses can be achieved
when the triangle degenerates into a line, i.e.\ when
$f_i (m_i \, \text{e}^{-\ci \, \phi_i})^d$ are real. In this case
\begin{align}
 m_3^d = \left| |f_1| m_1^d + |f_2| m_2^d \right| \text{ or }
 m_3^d =  \left| |f_1| m_1^d - |f_2| m_2^d \right| \;.
\end{align}
Replacing $m_3$ and $m_2$ with eq.~\eqref{eq:massspl} in this expression and solving
for $m_1$ we obtain, for instance, for $d=1$,
\begin{equation}
\label{eq:MassScaled1}
 \begin{split}
m_1^2 &= \frac{\Delta m_{21}^2 |f_2|^2 \left(|f_1|^2-|f_2|^2+1\right)+\Delta m_{31}^2   \left(|f_1|^2+|f_2|^2-1\right)}{|f_2|^4-2 \left(|f_1|^2+1\right) |f_2|^2+\left(|f_1|^2-1\right)^2} \\
   &\pm \frac{ 2 |f_1| |f_2| \sqrt{\Delta m_{31}^2 \left(\Delta m_{21}^2
   \left(|f_1|^2-1\right)+\Delta m_{31}^2\right)+\Delta m_{21}^2 |f_2|^2 (\Delta m_{21}^2-\Delta m_{31}^2)} }{|f_2|^4-2 \left(|f_1|^2+1\right) |f_2|^2+\left(|f_1|^2-1\right)^2}  \;,
   \end{split} 
\end{equation}
corresponding to the upper and lower bound for the mass scale.
As we will see later we will have here only $d=-1$ and $d=+1$.
However, the equivalent formula
for $d = -1$ is extremely lengthy and not insightful.
For this reason  we will present in the next section another approach leading to compact
but implicit expressions for all values of $d$.

The lower and upper bound on the lightest mass impacts the measurement on the sum
of the neutrino masses, $\sum m_i$, where the strongest current upper limit on the sum of the neutrino
masses as measured by Planck is $\sum m_i<0.12$ eV \cite{Aghanim:2018eyx}
as well as the kinematic neutrino mass
\begin{equation}
 m_\beta^2 = \sum_i  |U_{ei}|^2 m_i^2 \;,
\end{equation}
which can be measured with beta decay experiments like KATRIN  which provides a current limit of  $ m_\beta<1.1$ eV \cite{Aker:2019uuj} and is expected to reach a sensitivity of 0.2 eV in the future. Note that the kinematic neutrino mass also depends explicitly on the leptonic mixing angles  such that this observable gets also affected by the presence of a mixing sum rule.

\subsubsection{Neutrinoless double beta decay}
\label{sec:0vbb}

If neutrinos are Majorana particles, neutrinoless double beta decay is possible. In the minimal scheme
the decay rate is related to the parameter combination 
\begin{equation}
|m_{ee}| = \left|m_{1} U_{e1}^{2}+m_{2} U_{e2}^{2}+m_{3} U_{e3}^{2}\right|=\left| m_{1}c_{12}^{2}c_{13}^{2}\text{e}^{-\ci \phi_{1}}+m_{2}s_{12}^{2}c_{13}^{2}\text{e}^{-\ci \phi_{2}}+m_{3}s_{13}^{2}\text{e}^{-2 \ci \delta}\right| \;.
\label{eq:mee}
\end{equation}
Without a mass sum rule, this observable is being extremized when the
Majorana phases take the values
$2 \, \delta$ or $2 \, \delta + \pi$.
Then the factor $\exp(-2 \ci \delta)$ is a global unphysical factor. In the presence
of a mass sum rule this easy relation does not hold anymore since the Majorana phases are not independent
parameters.

In fact, we can use the mass sum rule to get expressions for $\phi_1$ and $\phi_2$.
We begin with $\phi_2$ and solve the mass sum rule for $m_{2} \, \text{e}^{-\ci\phi_{2}}$
such that
\begin{equation}
|m_{ee}| =\left| m_{1}c_{12}^{2}c_{13}^{2}\text{e}^{-\ci\phi_{1}}+ \frac{s_{12}^{2}c_{13}^{2}}{f_2^{1/d}} 
\left( - m_3^d - f_1 (m_1 \text{e}^{-\ci\phi_{1}})^d   \right)^{1/d}
+m_{3}s_{13}^{2}\text{e}^{-2 \ci \delta}\right| \;.
\label{eq:meeMSRgeneral}
\end{equation}
For instance, for $d=1$ this simplifies to
\begin{equation}
|m_{ee}| = \left| m_{1} \left(c_{12}^{2}c_{13}^{2} - \frac{f_1}{f_2}s_{12}^{2}c_{13}^{2}  \right)\text{e}^{-\ci\phi_{1}}+m_{3} \left(s_{13}^{2} \, \text{e}^{-2 \ci \delta} - \frac{s_{12}^{2}c_{13}^{2}}{f_2}  \right)\right| \;.
\label{eq:meed1}
\end{equation}
We see that $|m_{ee}|$ now only depends on $\phi_1$ since we eliminated $m_2$ and $\phi_2$.

To eliminate $\phi_1$ we solve the mass sum rule for  $f_2 (m_{2}\text{e}^{-\ci\phi_{2}})^d$
and  multiply it with its complex conjugate so that 
\begin{align}
 |f_2|^2 m_2^{2 \, d} &= |f_1|^2 m_1^{2 \, d} + m_3^{2 \, d} + 2 \, |f_1| \, m_1^d \, m_3^d \cos(\arg(f_1) - d \, \phi_1 ) \\
 \Leftrightarrow  \cos(\arg(f_1) - d \, \phi_1 ) &= \frac{|f_2|^2 m_2^{2 \, d} - |f_1|^2 m_1^{2 \, d} - m_3^{2 \, d}}{2 \, |f_1| \, m_1^d \, m_3^d} \\
  \Leftrightarrow   \phi_1 &= \frac{1}{d} \arg(f_1) \pm \frac{1}{d} \arccos\left(\frac{|f_2|^2 m_2^{2 \, d} - |f_1|^2 m_1^{2 \, d} - m_3^{2 \, d}}{2 \, |f_1| \, m_1^d \, m_3^d}\right) \;, \label{eq:phi1Solutions}
\end{align}
where we use the main branch of $\arccos(x) \in [0,\pi]$ and the $\pm$ then covers
the full range of $\phi_1$.
This expression for $\phi_1$ can be plugged into eq.~\eqref{eq:meeMSRgeneral}
eliminating all Majorana phases and reintroducing the dependence on $m_2$.
That is advantageous since for the masses (and mixing angles) we have experimental
information contrary to the Majorana phases
and $f_1$, $f_2$ and $d$ are given by the mass sum rule. The two solutions
of eq.~\eqref{eq:phi1Solutions} will enclose an allowed region in a plot $|m_{ee}|$
versus lightest neutrino mass as we will see later.

Furthermore, we can also determine the bounds on the mass scale from this equation
setting $\cos(\arg(f_1) - d \, \phi_1 ) = \pm 1$ and solving for the lightest neutrino mass.
This is in general a complicated formula, but no problem for common computer algebra
systems, in particular, after setting all parameters apart from the mass scale to numerical
values. For a given set of $f_1$, $f_2$, $d$ and mass squared differences it can also
happen that there are no real solutions for eq.~\eqref{eq:phi1Solutions} or that they
are outside of the experimentally allowed region excluding that particular parameter set
for the flavour model at hand.

\section{A detailed example}
\label{sec:exp}

As an explicit, detailed example for the new class of leptonic sum rules 
we analyse the model from \cite{Novichkov:2018yse} based on a modular $A_4$ symmetry.
Let us begin first with the parameter counting to identify how many sum rules
we expect. For the charged lepton sector two cases are present in this model,
which nevertheless
both just have three parameters describing the charged lepton masses which can
be made real by unphysical phase
transformations. The
difference between the two cases is a constant mixing matrix, to which we will
come back later.

Here we have to emphasize that the value of the moduli in this model
is fixed to certain values, which respect a residual symmetry. In the
charged lepton sector that is $\langle \tau_l \rangle = \ci \infty$ for
case I and $\langle \tau_l \rangle = -1/2 + \ci \sqrt{3}/2$
for case II. Both values leave a residual $Z_3$ symmetry intact
in the charged lepton sector.
In the neutrino sector the modulus is fixed to
$\langle \tau_\nu \rangle = \ci$ preserving a $Z_2$ symmetry.

The light neutrino mass matrix in this model has the structure
\begin{align}
 m_\nu &= a_1 \begin{pmatrix}
  2 & -1 & -1 \\
  -1 & 2 & -1 \\
  -1 & -1 & 2 
 \end{pmatrix}
 + a_2 \begin{pmatrix}
  1 & 0 & 0 \\
  0 & 0 & 1 \\
  0 & 1 & 0
 \end{pmatrix}
 + a_3 \begin{pmatrix}
  0 & 0 & 1 \\
  0 & 1 & 0 \\
  1 & 0 & 0
 \end{pmatrix} \nonumber\\
 &= \frac{c}{\sqrt{3}} \begin{pmatrix}
  1 & 0 & 2 \\
  0 & 2 & 1 \\
  2 & 1 & 0
 \end{pmatrix}
 + \frac{a c}{3} \begin{pmatrix}
  5 & -1 & -1 \\
  -1 & 2 & 2\\
  -1 & 2 & 2
 \end{pmatrix}
 + \frac{b c}{3} \begin{pmatrix}
   1 & 1 & 1 \\
   1 & -2 & 4\\
   1 & 4 & -2
 \end{pmatrix} \;.
 \label{eq:MnuExample}
\end{align}
Here we have three complex parameters and we can absorb one unphysical
phase (the other two phases are relative) describing three masses,
three mixing angles and three CPV phases.
Therefore we expect in total
four predictions which in the standard parametrisation are relations
between observables. Since we want to write down a neutrino mass sum rule,
which gives two relations, 
we can find two more relations between the mixing angles and the Dirac CP
phase.

Let us begin with case I in this model.
The neutrino mass matrix can be diagonalised by
\begin{align}
\label{eq:TBM13}
 m_\nu^{\text{diag}} = U_{13}(\theta,\phi)^T U_{\text{TBM}}^T  m_\nu U_{\text{TBM}}  U_{13}(\theta,\phi) \;
\end{align}
where we use the following phase convention for
the so-called tri-bimaximal (TBM) mixing matrix~\cite{Harrison:2002er}
\begin{align}
U_{\text{TBM}}=\begin{pmatrix}
2/\sqrt{6}& 1/\sqrt{3}& 0\\
-1/\sqrt{6}& 1/\sqrt{3}& -1/\sqrt{2}\\
 -1/\sqrt{6}& 1/\sqrt{3}& 1/\sqrt{2}
\end{pmatrix} \;.
\end{align}
The matrix $U_{13}$ is a unitary rotation matrix
\begin{equation}
 U_{13}(\theta,\phi) = \begin{pmatrix}
  \cos \theta & 0 & \text{e}^{-\ci \phi} \sin \theta \\
  0 & 1 & 0 \\
  -\text{e}^{\ci \phi} \sin \theta & 0 & \cos \theta
 \end{pmatrix} \;,
\end{equation}
depending on the angle $\theta$ and the phase $\phi$ and
which accounts for non-zero $\theta_{13}$.
Note that we use a different sign convention for $\phi$
compared to \cite{Novichkov:2018yse}.

The fact that the neutrino mixing matrix has such a simple structure
is a consequence of the residual symmetry. The columns of the TBM matrix
are indeed eigenvectors of the preserved $Z_2$ generator
\begin{equation}
 S = \frac{1}{3} \begin{pmatrix}
   -1 &  2 &  2 \\
    2 & -1 &  2 \\
    2 &  2 & -1
 \end{pmatrix} \;,
\end{equation}
which is a symmetry of the mass matrix eq.~\eqref{eq:MnuExample}, cf.~\cite{Novichkov:2018yse}.
We will see such patterns again in the other models we consider.
 
Then the three neutrino masses are (see \cite{Novichkov:2018yse}) 
\begin{align}
 m_1 \, \text{e}^{-\ci \varphi_1  } &= c \left( z - \frac{1}{\sin 2\theta} \right) \, \text{e}^{\ci  \phi}  \;, \\
 m_2 \, \text{e}^{-\ci \varphi_2  } &= c \left( \sqrt{3} + \ci z \sin \phi - \cot 2\theta \cos \phi \right) \;, \\
 m_3 \, \text{e}^{-\ci \varphi_3  }  &= c \left( z + \frac{1}{\sin 2\theta} \right) \, \text{e}^{-\ci \phi} \;, 
\end{align}
where $z = a \, \, \text{e}^{-\ci \phi} - b \, \text{e}^{\ci \phi}$. Note that
we use here $\varphi_i$ instead of $\phi_i$ since on the left-hand side
we do not have yet fixed a convention for the Majorana phases. 

The coefficients of the mass sum rule, $f_1$ and $f_2$, are functions of $\theta$ and $\phi$
\begin{align}
 f_1 &= - \text{e}^{-2 \ci \phi} - \ci \text{e}^{-\ci \phi} f_2 \sin \phi \;, \nonumber\\
  &= - \text{e}^{-2 \ci \phi}  \frac{\sqrt{3} \sin(2\theta) - \cos \phi \cos(2\theta) - \ci \sin \phi}{\sqrt{3} \sin(2\theta) - \cos \phi \cos(2\theta) + \ci \sin \phi} \\
 f_2 &= - \text{e}^{-\ci \phi} \frac{ 2  }{ \sqrt{3} \sin(2\theta) - \cos \phi \cos(2\theta) + \ci \sin \phi } \;.
 \label{eq:f1f2}
\end{align}
It is obvious from these expressions that $|f_1| = 1$ while $|f_2|$ is non-trivial.
In \cite{Novichkov:2018yse} the authors did not calculate explicitly
a mass sum rule. This result is hence new. We will derive five different mass sum
rules in this paper from which four had not been derived explicitly in the original model.

From the mixing sum rules of the model \cite{Novichkov:2018yse} we additionally obtain
the following predictions for the mixing parameters as a function of $\theta$ and $\phi$
(these relations were found before in \cite{Grimus:2008tt,Petcov:2017ggy})
\begin{align}
\label{eq:sr1_mix1}
 \sin^2 \theta_{12}(\theta) &= \frac{1}{3 - 2 \sin^2 \theta} \;, \\
 \label{eq:sr1_mix2}
 \sin^2 \theta_{13}(\theta) &= \frac{2}{3} \sin^2 \theta \;, \\
 \label{eq:sr1_mix3}
 \sin^2 \theta_{23}(\theta,\phi) &= \frac{1}{2} + \frac{\sin \theta_{13}(\theta) }{2} \frac{\sqrt{2 - 3    \sin^2 \theta_{13}(\theta) }}{1 - \sin^2 \theta_{13}(\theta) } \cos \phi \;, \\
 \label{eq:sr1_mix4}
 \delta(\theta,\phi) &= \arcsin\left(- \frac{\sin \phi}{\sin 2 \theta_{23}(\theta,\phi) } \right) \;,
\end{align}
where we did not write the dependence on $\theta$ and $\phi$ completely explicit
in the last two equations for better readability. These four equations allow us to fix
$\theta$ and $\phi$ and then give two additional relations (predictions).
These formulas together with the above coefficients for a mass sum rule form
what we will call sum rule 1 (SR 1) case I.

In order to obtain the experimentally allowed ranges for the model parameters  $\theta$ and $\phi$
we confront eqs.~\eqref{eq:sr1_mix1}-\eqref{eq:sr1_mix4} with global neutrino oscillation data
where we 
use the one-dimensional $\chi^2$-profiles for these parameters provided
by {\tt nu-fit}~v5.0~\cite{Esteban:2020cvm}
to determine the total $\chi^2(\theta,\phi)$
\begin{align}
  \chi^2(\theta,\phi)=&\left(\frac{\sin^2\theta_{12}(\theta,\phi)-\sin^2\theta_{12}}{\sigma(\sin^2\theta_{12})}\right)^2+\left(\frac{\sin^2\theta_{13}(\theta,\phi)-\sin^2\theta_{13}}{\sigma(\sin^2\theta_{13})}\right)^2
 \nonumber \\&+\left(\frac{\sin^2\theta_{23}(\theta,\phi)-\sin^2\theta_{23}}{\sigma(\sin^2\theta_{23})}\right)^2+
  \left(\frac{\delta(\theta,\phi)-\delta}{\sigma(\delta)}\right)^2 \;.
\end{align}

The minimal $\chi^2$ for case I in this model is $\chi_{\text{min}}^2=8.6$ for NO,
while the minimal $\chi^2$ for IO is 19.7, higher than for NO due to
the current mild preference of the global fit for NO which leads to a minimal $\chi^2$
for IO of 10.8 in the absence of sum rules.
The $\chi^2$ difference between the case with
and without sum rules is similar in both cases, 8.6 and 8.9 respectively.

For case II the charged
lepton sector is non-diagonal such that a contribution to the PMNS matrix from the mixing
matrix which diagonalises the charged lepton mass matrix arises. The relation between
the PMNS matrix in case I and case II is, cf.~\cite{Novichkov:2018yse},
\begin{align}
U_{\text{PMNS}}^{\text{II}}=\begin{pmatrix}
-1&0&0\\
0&\text{e}^{\ci\pi/3}&0\\
0&0&\text{e}^{-\ci\pi/3}\\
\end{pmatrix} 
U_{\text{PMNS}}^{\text{I}}
\begin{pmatrix}
\text{e}^{- \ci(\phi+\pi/2)}&0&0\\
0&1&0\\
0&0&\text{e}^{\ci(\phi - \pi/2)}\\
\end{pmatrix}~.
\end{align}
The light neutrino masses of 
case II are related to case I via
\begin{align}
m_1^{(\text{II})}  \text{e}^{- \ci \phi_1^{(\text{II})}} &=  m_1^{(\text{I})}  \text{e}^{- \ci \phi_1^{(\text{I})}} \text{e}^{4 \ci  \phi} \;,\\
m_2^{(\text{II})}  \text{e}^{- \ci \phi_2^{(\text{II})}} &=  m_2^{(\text{I})}  \text{e}^{- \ci \phi_2^{(\text{I})}} (- \text{e}^{-2 \ci \phi  } ) \;,
\end{align}
while the mixing angles do not change and we can use the fit result
for $\theta$ and $\phi$ in both cases. The coefficients in the mass sum rule
change nevertheless
\begin{align}
f_1^{(\text{II})}  &=  f_1^{(\text{I})} \text{e}^{-4 \ci \phi} \;, \\ 
f_2^{(\text{II})}  &=  - f_2^{(\text{I})} \text{e}^{2 \ci \phi } \;.
\end{align}
These changed coefficients together with the mixing sum rules
in eqs.~\eqref{eq:sr1_mix1}-\eqref{eq:sr1_mix4} form SR~1 case II.

\begin{figure}
    \centering
    \includegraphics[width=0.45\linewidth]{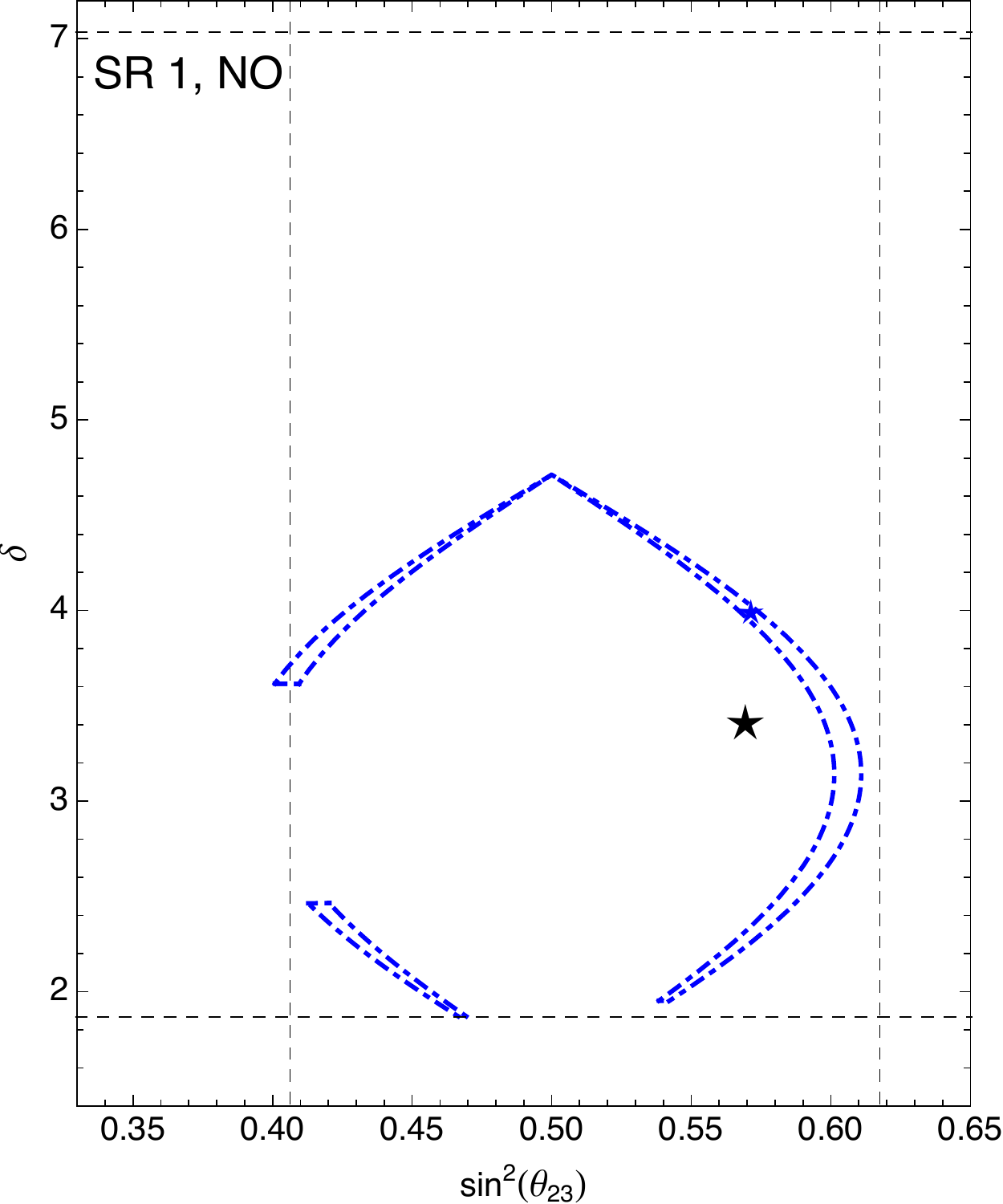}
    \includegraphics[width=0.45\linewidth]{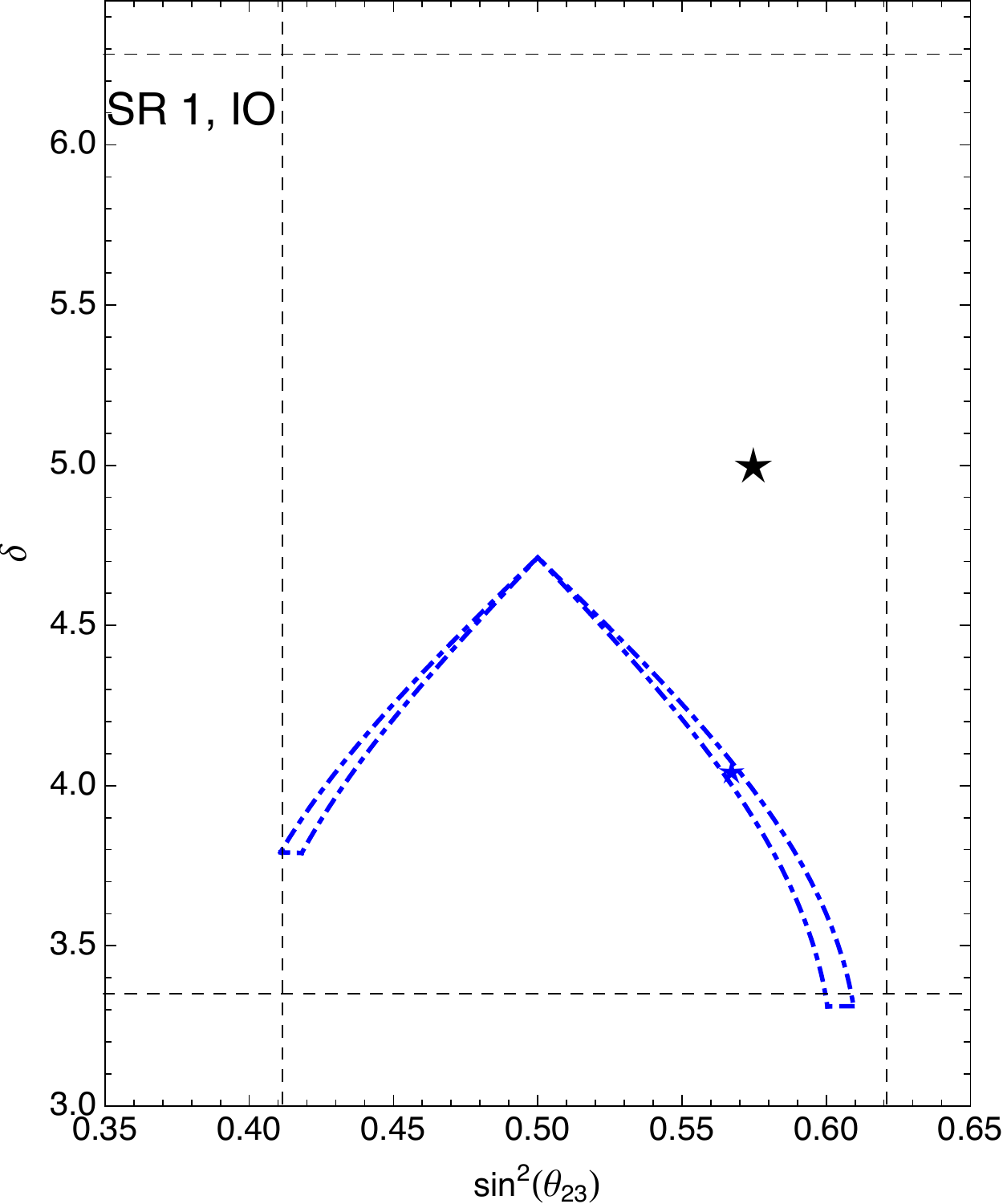}
    \caption{Correlations between $\delta$ and $\theta_{23}$ in blue for SR 1 for
      NO (left) and IO (right). The blue lines have been obtained by varying the model
      parameters $\theta$ and $\phi$ in their $3\sigma$ ranges. The dashed lines show the
      experimental $3\sigma$ ranges from  {\tt nu-fit 5.0}. The blue (black) star represents
      the best fit value with (without) the presence of SR 1.
      \label{fig:th23deltaSR1}}
\end{figure}

Before turning to the predictions of the mass sum rule we discuss the predictions 
resulting from the mixing sum rules, eqs.~\eqref{eq:sr1_mix1}-\eqref{eq:sr1_mix4}.
One obvious immediate consequence is that $\sin^2 \theta_{13}$ and $\sin^2 \theta_{12}$
are strongly correlated, i.e.
\begin{equation}
 3 \sin^2 \theta_{12} = \frac{1}{1 - \sin^2 \theta_{13}} \;.
\end{equation}
Since $\theta_{13}$ is the best-known mixing angle this predicts 
a value of $\theta_{12}$ around $35.7^\circ$ close to the experimental
upper $3\sigma$ bound on $\theta_{12}$.
More interesting though is the correlation between the rather badly known
$\theta_{23}$ and the CPV phase $\delta$ depicted in
Fig.~\ref{fig:th23deltaSR1}. The blue area encloses the region where
we vary $\theta$ and $\phi$ within its 3$\sigma$ ranges for NO and IO,
respectively, the blue star denoting the best fit point in the model.
The dashed rectangle encloses the
3$\sigma$ ranges and the black star stands for the best fit point
of {\tt nu-fit 5.0}.
Compared to the unconstrained fit the allowed regions
are much smaller. We also see that the best fit  of the model prefers $\theta_{23}$ 
in the upper octant. For most of the possible values of $\delta$ two degenerate
solutions for $\theta_{23}$ in the upper and lower octant are possible due to the fact
that eq.~\eqref{eq:sr1_mix4} is symmetric around $\theta_{23}=45^\circ$ while
maximal atmospheric mixing is only possible if $\delta=3\pi/2$
(which has also been noticed by the authors of  \cite{Novichkov:2018yse}).
This is obvious from eq.~\eqref{eq:sr1_mix4} which states that for
maximal $\theta_{23}$ the Dirac CP phase $\delta=-\phi$,
independent of the value of $\theta$.
Maximal $\theta_{23}$ can only be achieved if the second term
in eq.~\eqref{eq:sr1_mix3} is zero, that is $\phi = \pm \pi/2$ such that $\delta = \mp \pi/2$.
However, $\delta=\pi/2$  is currently disfavoured
by more than $3\sigma$ by the global oscillation fit such that we only find the
solution $3\pi/2$ in our model.

\begin{figure}
    \centering
    \includegraphics[width=0.45\linewidth]{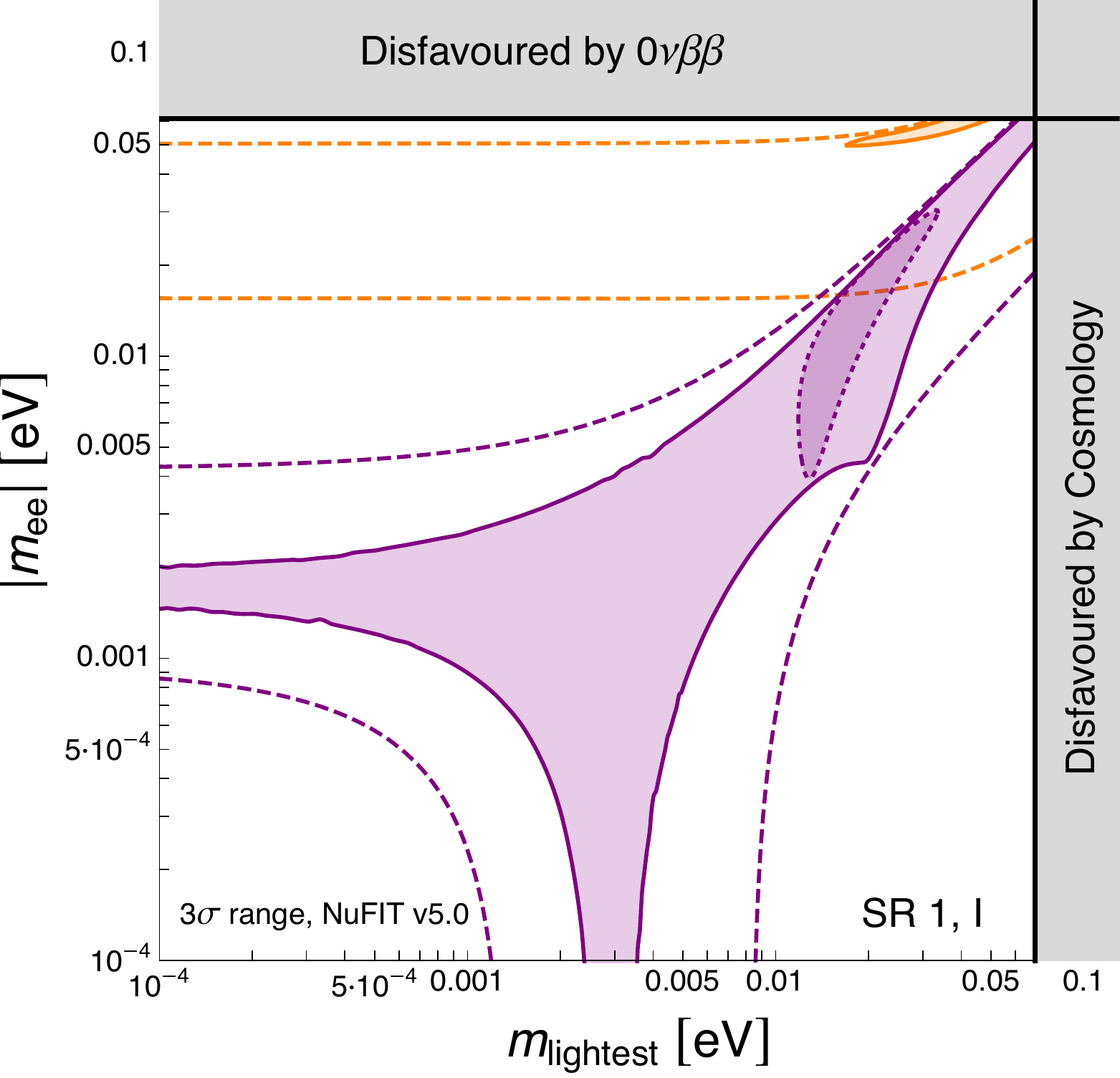}  \hspace{0.2cm}
    \includegraphics[width=0.45\linewidth]{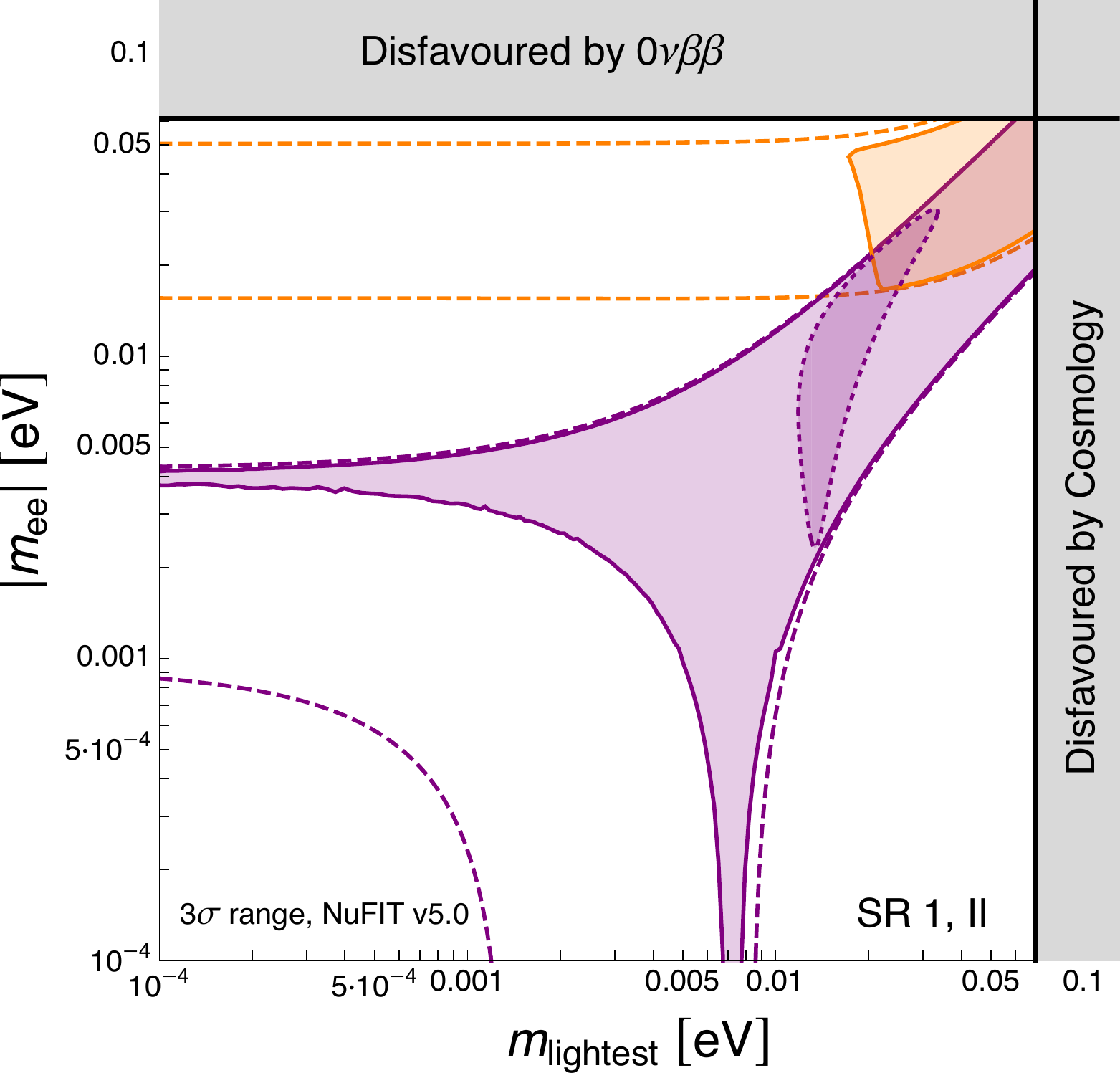}
    \caption{Allowed ranges for $|m_{ee}|$ for SR 1 in case I and case II.
    The purple/orange dashed region is the $3\sigma$ allowed region without
    sum rules for NO/IO. The lightly (dark) shaded purple/orange regions
    is the $3\sigma$ (best fit) allowed region for NO/IO. For more details,
    see main text.
    \label{fig:meeSR1}}
\end{figure}

Turning now to the predictions of the mass sum rule.
For the best fit values of $\theta$ and $\phi$ the coefficients of the mass sum rules
are for NO in case I
\begin{align}
 f_1^{(\text{I})} \approx -0.11 - 0.99 \ci \text{ and } f_2^{(\text{I})} \approx 2.02+ 1.80\ci \;,
\end{align}
and for IO
\begin{align}
 f_1^{(\text{I})} \approx -0.11 - 0.99 \ci \text{ and } f_2^{(\text{I})} \approx 1.92+ 1.72\ci \;.
\end{align}
For case II we find
\begin{align}
f_1^{(\text{II})} \approx 0.29 + 0.96   \ci \text{ and } f_2^{(\text{II})} \approx 1.98 - 1.85  \ci \;,
\end{align}
for NO and for IO
\begin{align}
f_1^{(\text{II})} \approx 0.99+0.09  \ci \text{ and } f_2^{(\text{II})} \approx 2.58+0.11  \ci \;.
\end{align}
The results for  $|m_{ee}|$ for NO and IO and case I and II
are shown in fig.~\ref{fig:meeSR1}.
We see that SR~1 puts strong constraints on neutrinoless double beta
decay, especially for IO where we also find that for the best fit values of
$\theta$ and  $\phi$ the SR cannot be fulfilled such that we only obtain a $3\sigma$
range in IO.  We have seen this constraining power already 
for the previous mass sum rules, cf.~\cite{King:2013psa, Gehrlein:2016wlc,Gehrlein:2016fms}.

There is nevertheless, one major difference in this model compared to the previous cases.
For NO there is no lower bound on the mass scale which has been already noted in
\cite{Novichkov:2018yse}, which can be easily understood in our formalism.
From the mass sum rule the absence of a lower bound, i.e., $m_1 = 0$, implies
\begin{align}
 1 \ll \frac{m_3}{m_2} =  \frac{\sqrt{\Delta m_{31}^2}}{\sqrt{\Delta m_{21}^2}}
  &= |f_2| = \frac{2}{| \sqrt{3} \sin(2\theta) - \cos \phi \cos(2\theta) - \ci \sin \phi| } \;.  
\end{align}
So if the denominator is close to zero this can be fulfilled which is indeed the case
in our 3$\sigma$ region.

Similarly, there is also no upper bound on the mass scale for both NO and IO and we can
understand this as well. An upper bound on the mass scale implies that there
is a mass scale for which
\begin{align}
 1 &< |\cos(\arg(f_1) - \phi_1 )| = \left| \frac{|f_2|^2 m_2^{2} - |f_1|^2 m_1^{2} - m_3^{2}}{2 \, |f_1| \, m_1 \, m_3} \right| \;,
\end{align} 
where we have used that $d = 1$. If there is no mass bound this implies that
we can go to arbitrary large masses, where the mass splittings are negligible
and still find a valid value for $\phi_1$.
In that limit $m_1 \approx m_2 \approx m_3$ and
\begin{align}
 |\cos(\arg(f_1) - \phi_1 )| \approx \left| \frac{|f_2|^2 - |f_1|^2 - 1}{2 \, |f_1| } \right|
 = \left|1 -  \frac{|f_2|^2}{2} \right| \;,
\end{align} 
where we have used $|f_1|=1$. In the allowed $3\sigma$ regions for both
orderings $|f_2|$ can be smaller than two and hence there is no upper
bound on the mass scale.

\begin{figure}
    \centering
    \includegraphics[width=0.45\linewidth]{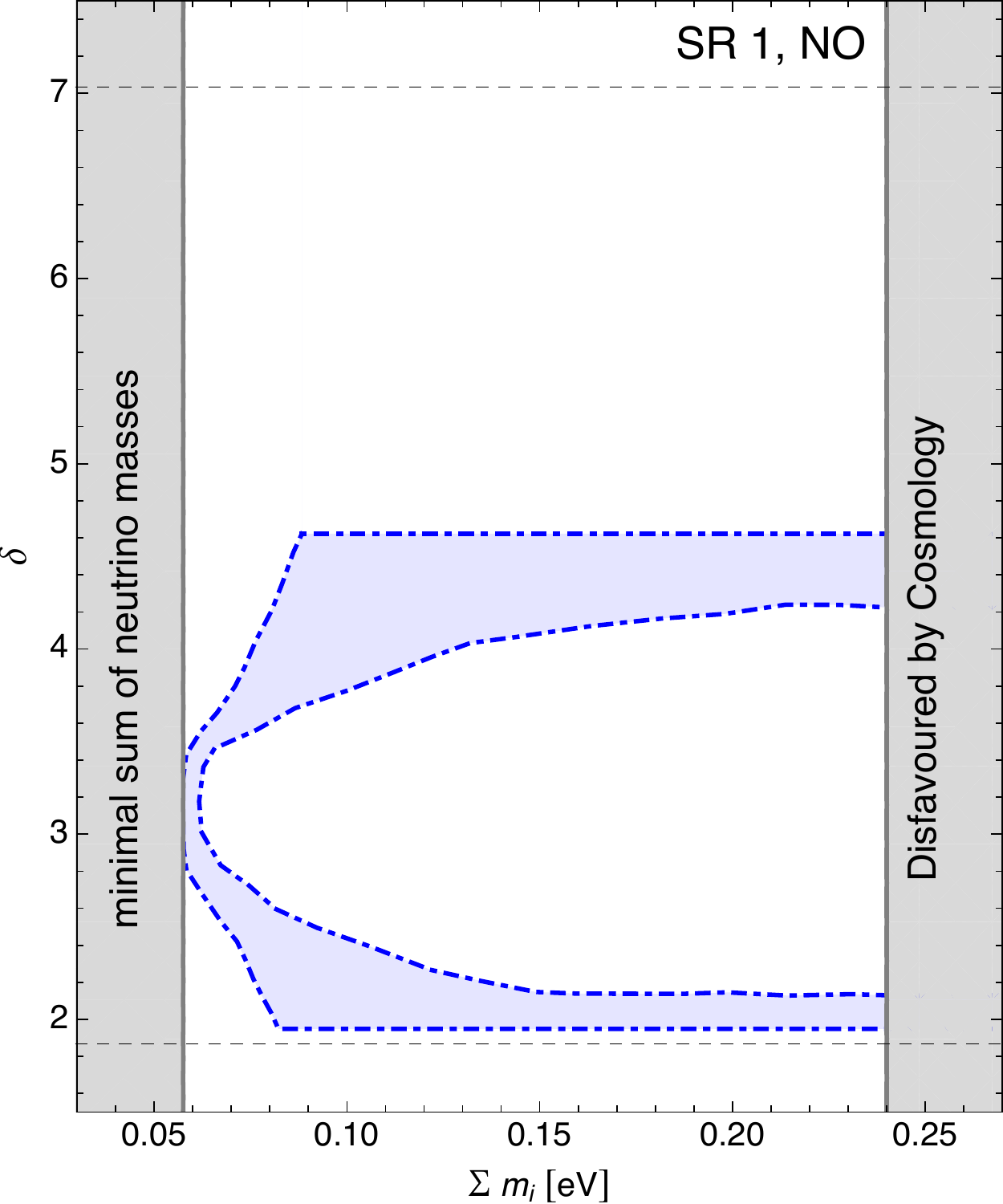} \hspace{0.2cm}
    \includegraphics[width=0.45\linewidth]{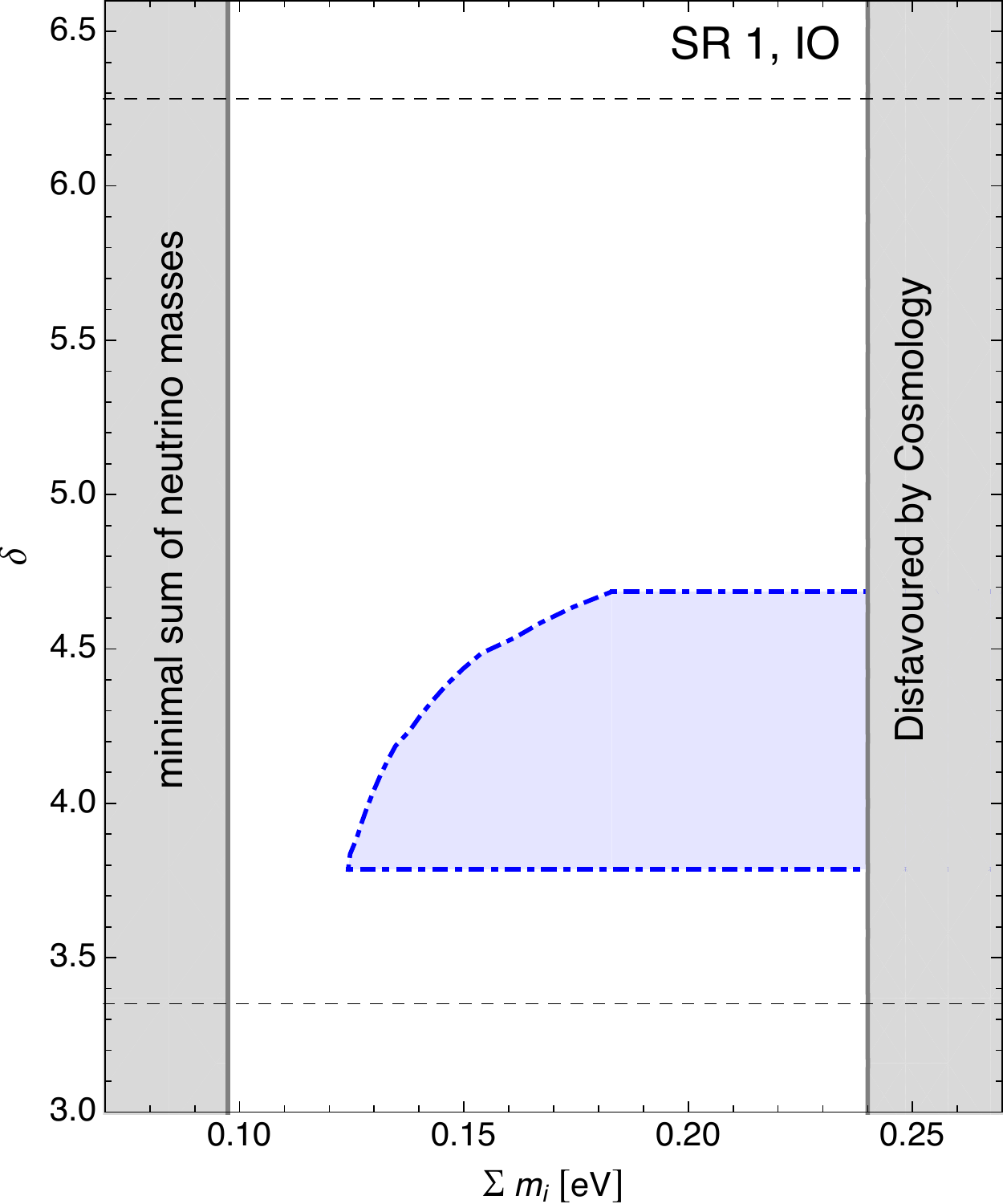}\\[0.5cm]
    \includegraphics[width=0.45\linewidth]{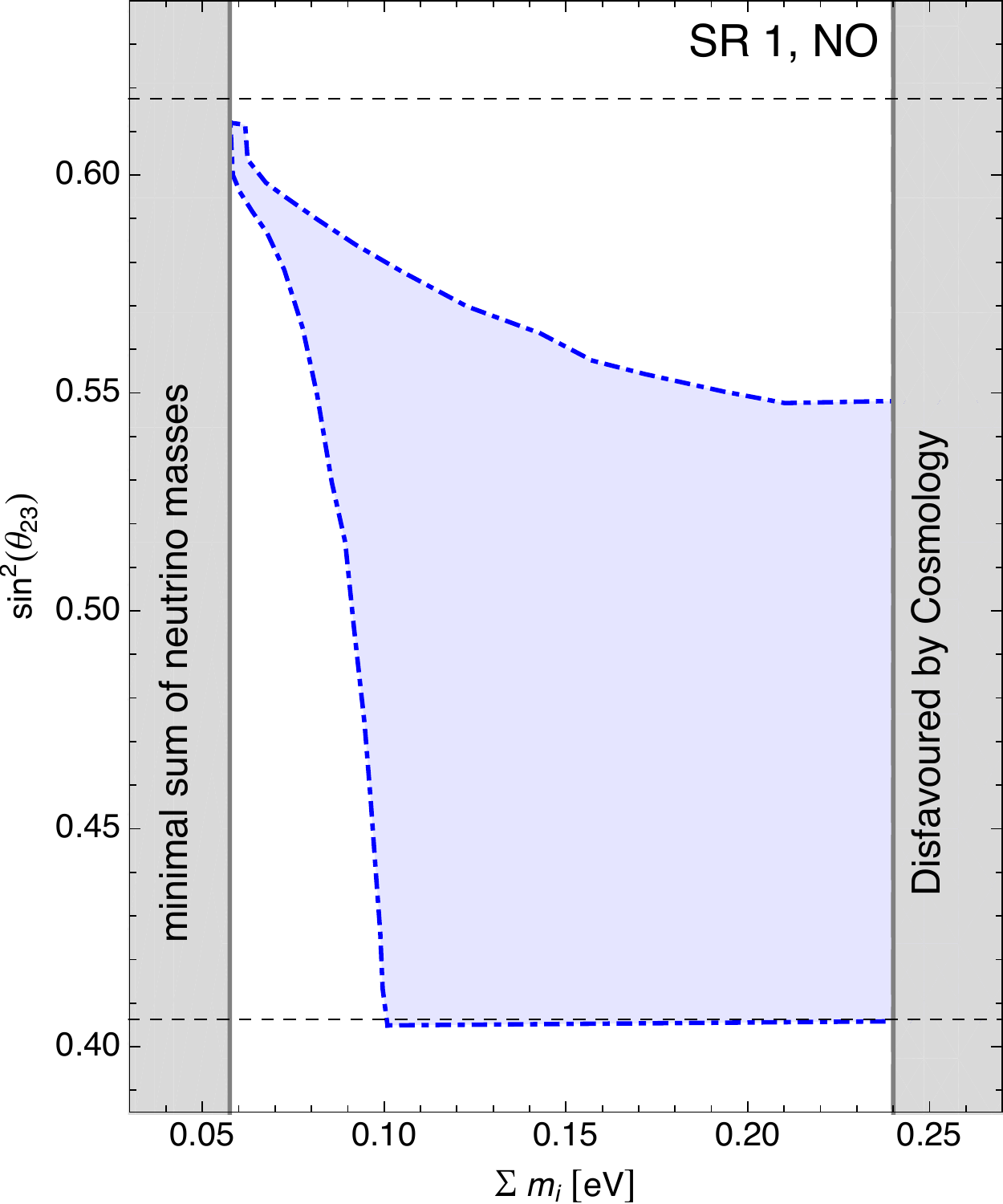} \hspace{0.2cm}
    \includegraphics[width=0.45\linewidth]{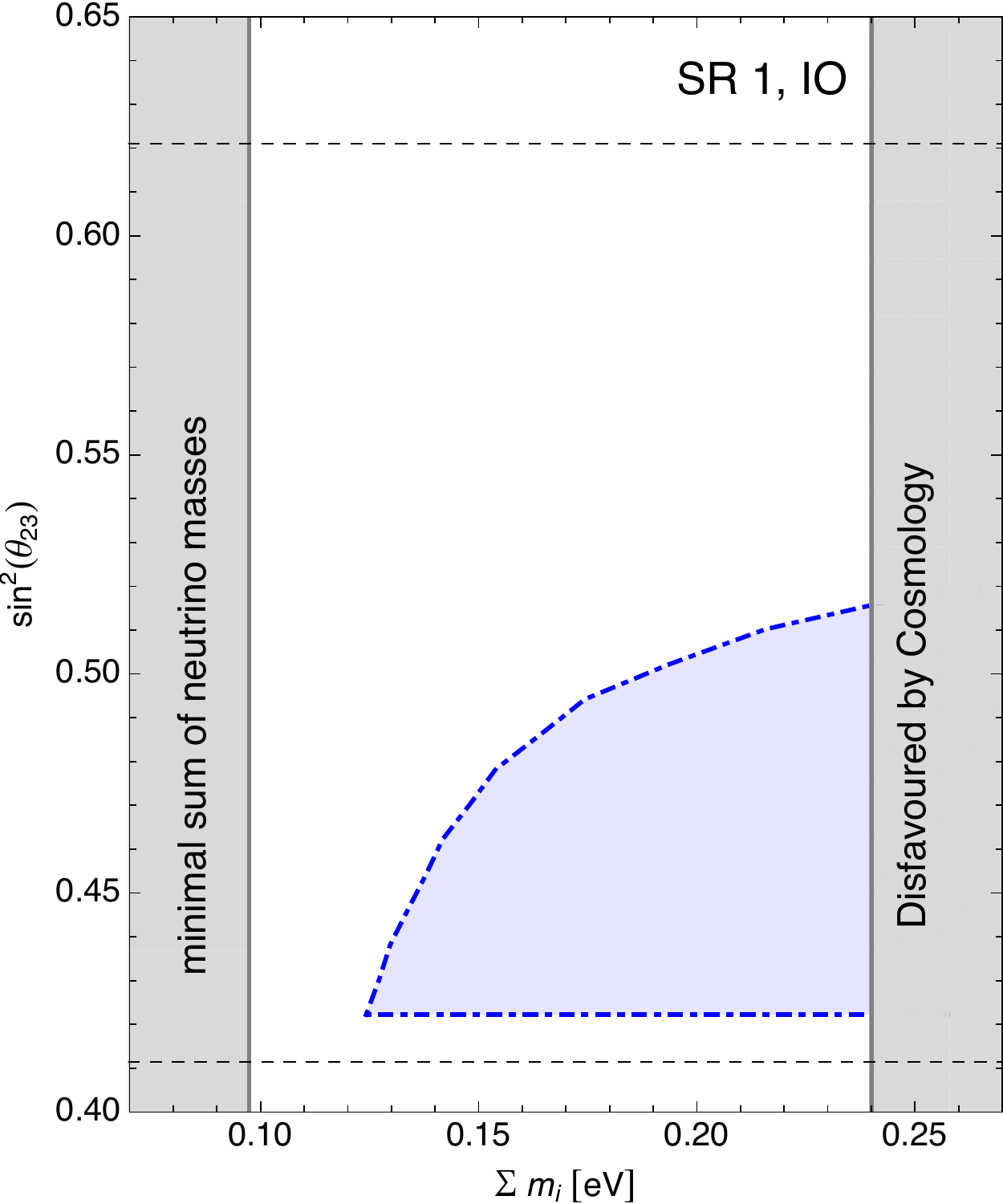}
    \caption{Correlation between $\delta$ (upper plots) or $\sin^2\theta_{23}$ (lower plots)
      and the sum of the neutrino masses in SR 1 for NO (left plots) and IO (right plots)
      for $\theta,~\phi$ and the mass splittings in their $3\sigma$ ranges.
      The black dashed lines show the $3\sigma$ allowed region without
      sum rules for NO/IO. The gray exclusion regions show the minimal value of the sum of the
      neutrino masses from oscillation experiments and the maximal value of the sum of
      the neutrino masses from cosmology.
      \label{fig:summSR1}}
\end{figure}

Finally, the true power of modular flavour models shows
if we consider the correlations between mixing parameters and neutrino masses.
In Fig.~\ref{fig:summSR1} we display the correlation between
$\theta_{23}$ and the sum of neutrino masses and $\delta$ and the sum of neutrino
masses for NO and IO respectively. In particular, the minimal mass scale
depends on the values of $\theta_{23}$ and $\delta$ which has also been
seen in~\cite{Novichkov:2018yse}. This model can therefore
be constrained and potentially excluded by all(!) experiments measuring
neutrino parameters like oscillation experiments but also experiments which
are sensitive to the neutrino mass scale.

\section{More models with sum rules: Formulas}
\label{sec:models}

In the following we want to discuss more examples for flavour models
based on modular symmetries which exhibit sum rules. There are many potential
candidates but we will not consider models that have already been excluded or that have more
parameters than observables in the neutrino sector. As the maximal number of sum
rules is achieved when residual symmetries are present we will restrict ourselves
to these cases.
This means we want to restrict ourselves further on models where the values of the moduli
are fixed on symmetry values. This allows to write down somewhat involved, but explicit
expressions for mass sum rules as we have seen in Sec.~\ref{sec:exp}.
To keep the number of parameters in the neutrino sector manageable we also do not
consider radiative models here.
In the general case, mass and/or mixing sum rules can still exist
if there are less parameters than observables, but it can be difficult
and not insightful to write them down. The existence of mass sum rules in particular is also suggested
by the fact that many models with modular flavour symmetries have a lower bound on the
neutrino mass scale which is a generic prediction of mass sum rules.

It is also interesting to note that in many cases the best fit value for
the expectation value of the modulus, $\langle\tau\rangle$,
 is near a symmetry point, see, for instance,
\cite{Novichkov:2020eep, Okada:2020ukr}, where formalisms have been developed which allows to 
expand around symmetry points.

This section provides the analytical expressions for the sum rules, in the next
section we show a summary of the numerical results for them
and compare them to our explicit example from the previous section.
The models in this section have no particular order.

\subsection{A model with two modular $\boldsymbol{S_4}$ groups}
\label{sec:S4Sq}

In  \cite{King:2019vhv}  a model with two modular $S_4$ groups was discussed.
The modulus value $\langle \tau_l \rangle$ in the charged lepton sector is fixed
to $\omega  = \exp(2\, \pi \ci/3)$ such that there is
a residual $Z_3^T$ symmetry making the Yukawa matrix diagonal.
In the neutrino sector  $\langle \tau_\nu \rangle = (\ci-1)/2$ such that a
$Z_2^{SU}$ symmetry is preserved with
\begin{equation}
 SU = \frac{1}{3} \begin{pmatrix}
-1 & 2 & 2 \\
2 & 2 & -1 \\
2 & -1 & 2
\end{pmatrix} \;.
\end{equation}
Due to this preserved generator which has the TBM matrix columns as
eigenvector the neutrino mixing matrix has the trimaximal TM$_1$ structure
\cite{Varzielas:2012pa, Luhn:2013lkn}
\begin{align}
U_{\text{TM}_1} = U_{\text{TBM}} U_{23}=
U_{\text{TBM}}\begin{pmatrix}
1&0&0\\
0&\cos\theta &\sin\theta \, \text{e}^{-\ci \phi}\\
0&-\sin\theta \, \text{e}^{\ci \phi}&\cos\theta \\
\end{pmatrix} 
\end{align}
with the same TBM matrix as in sec.~\ref{sec:exp} corrected by a 2-3 rotation.
Note that we deviate here from the conventions of the 2-3 rotation
in \cite{King:2019vhv}
for consistency with the rest of our paper.

The right-handed neutrino mass matrix in this model is
\begin{align}
M_{R}&= a\begin{pmatrix}
1 & 0 & 0 \\
0 & 0 & 1 \\
0 & 1 & 0
\end{pmatrix}+b\begin{pmatrix}
0 & 1 & 1 \\
1 & 1 & 0 \\
1 & 0 & 1
\end{pmatrix} 
+c \sqrt{2}\begin{pmatrix}
2 & -1 & -1 \\
-1 & 2 & -1 \\
-1 & -1 & 2
\end{pmatrix}-c \sqrt{3}\begin{pmatrix}
0 & 1 & -1 \\
1 & 2 & 0 \\
-1 & 0 & -2
\end{pmatrix}
\end{align}
and the Dirac neutrino mass matrix is
\begin{equation}
 M_D = y_D v_u \left(
\begin{array}{ccc}
 1 & 0 & 0 \\
 0 & 0 & 1 \\
 0 & 1 & 0 \\
\end{array}
\right) \equiv y_D v_u P_{23} \;,
\end{equation}
where we have defined the 2-3 permutation matrix $P_{23}$.
We first diagonalise the right-handed neutrino mass matrix
\begin{equation}
 U_{\text{TM}_1}^T M_R U_{\text{TM}_1} = \diag(\tilde{M}_1,\tilde{M}_2,\tilde{M}_3) \;,
\end{equation}
where we eliminate $b$ using
\begin{equation}
 b = \frac{\text{e}^{2 \ci \phi} \left(a-3 \sqrt{2} c\right)+a+6 \sqrt{2} \, c \, \text{e}^{\ci \phi} \cot (2
   \theta)}{-2+\text{e}^{2 \ci \phi}} \;.
\end{equation}
The parameters $\theta$ and $\phi$ can be determined
from oscillation data as we will discuss later.

With the seesaw formula
\begin{align}
 m_\nu = - M_D^T M_R M_D = - y_D^2 v_u^2 P_{23}  U_{\text{TM}_1} \diag(\tilde{M}_1,\tilde{M}_2,\tilde{M}_3)^{-1} U_{\text{TM}_1}^T P_{23}^T  
\end{align}
it is straightforward to write down the PMNS mixing matrix 
\begin{equation}
 U_{\text{PMNS}} = P_{23} U_{\text{TBM}}^* U_{23}^* \Gamma_i \;,
\end{equation}
where $\Gamma_i$ is a diagonal phase matrix which contains a global unphysical phase
and the two physical Majorana phases. It is easy to derive that the light neutrino masses
are given as
\begin{align}
 -y_D^2 v_u^2 \tilde{m}_1^{-1} &=  \tilde{M}_1 =c \left(-\frac{6 \sqrt{2} \text{e}^{\ci \phi} \cot (2 \theta)}{-2+\text{e}^{2 \ci \phi}}+\frac{6 \sqrt{2}}{-2 + \text{e}^{2 \ci \phi}}+6
   \sqrt{2}\right)-\frac{3 \, a}{-2 + \text{e}^{2 \ci \phi}} \;, \\
 -y_D^2 v_u^2 \tilde{m}_2^{-1} &=  \tilde{M}_2 = c \frac{ \left(-3 \sqrt{2} \text{e}^{\ci \phi} \tan (\theta) \left(-2 \cot
   ^2(\theta)+\text{e}^{2 \ci \phi}\right)-6 \sqrt{2} \text{e}^{2 \ci \phi}\right)}{-2 + \text{e}^{2 \ci \phi}} + \frac{3 \, a \, \text{e}^{2 \ci \phi}}{-2+\text{e}^{2 \ci \phi}}\;, \\
 - y_D^2 v_u^2 \tilde{m}_3^{-1} &=  \tilde{M}_3 = c \frac{3 \left(-2 \sqrt{2}+\sqrt{2} \text{e}^{- \ci \phi} \tan (\theta) \left(-2+ \text{e}^{2 \ci
   \phi} \cot^2(\theta)\right)\right)}{-2+\text{e}^{2 \ci \phi}} +\frac{3 \, a}{-2+\text{e}^{2 \ci \phi}} \;.
\end{align}

From that we can determine the coefficients for the mass sum rule
\begin{align}
 f_1 &= \frac{1}{\cos^2 \theta -  \text{e}^{\ci \phi} \sin (2 \theta)} \;, \\
 f_2 &=  -\frac{\tan \theta + 2 \, \text{e}^{\ci \phi}}{2 \,  \text{e}^{3  \ci \phi} - \text{e}^{2 \ci \phi} \cot (\theta)} \;, \\
   d &= -1 \;.
\end{align}
From these equations it is noteworthy that the right-handed neutrino masses fulfill a
mass sum rule with the same $f_1$ and $f_2$ but $d = +1$.
The authors of \cite{King:2019vhv} also derive a mass sum rule which leads to the
same predictions as our sum rule here. We only differ in conventions. The
advantage of our approach is that we can immediately compare the result to
other sum rules derived here or already present in the literature like in \cite{Altarelli:2008bg, Hirsch:2008rp, Bazzocchi:2009da, Altarelli:2009kr, Chen:2009um, Barry:2010yk, Dorame:2011eb, King:2013psa, Agostini:2015dna, Gehrlein:2015ena, Gehrlein:2016wlc}.

From the PMNS matrix we can also easily determine the relations between $\theta$, $\phi$
and the mixing angles in the standard convention 
\begin{align}
\sin\theta_{13}&= |(U_{\text{PMNS}})_{13}| =\frac{\sin\theta}{\sqrt{3}}~,\\
\tan\theta_{12}&= \frac{|(U_{\text{PMNS}})_{12}|}{|(U_{\text{PMNS}})_{11}|}  = \frac{\cos\theta}{\sqrt{2}}~,\\
\tan \theta_{23}&= \frac{|(U_{\text{PMNS}})_{23}|}{|(U_{\text{PMNS}})_{33}|} =\left|\frac{\cos\theta+\sqrt{\frac{2}{3}}\text{e}^{\ci \phi} \sin\theta}{\cos\theta-\sqrt{\frac{2}{3}}\text{e}^{\ci \phi}\sin\theta}\right|~.
\end{align}
To derive an expression for the Dirac CP phase we first calculate
\begin{align}
 \Delta &= \frac{(U_{\text{PMNS}})_{11}^* (U_{\text{PMNS}})_{13} (U_{\text{PMNS}})_{31} (U_{\text{PMNS}})_{33}^*}{ s_{12} s_{23} c_{12} c_{13}^2 c_{23} s_{13} } + \frac{c_{12} c_{23} s_{13}}{s_{12} s_{23}} 
\end{align}
which is basis independent and in the standard parametrisation just  $\exp(- \ci \delta)$. Therefore
\begin{equation}
 \tan \delta = - \frac{\Im(\Delta)}{\Re(\Delta)} = - \frac{ 5 + \cos(2 \theta)}{1+ 5 \cos (2 \theta) } \tan\phi \;.
\end{equation}
As we can see $\theta_{13}$ is determined by $\theta$ which implies $\theta\ll 1$.
In this case $\delta\approx -\phi$ such that maximal $\theta_{23}$ is only possible
for $\phi= -\delta=\pm\pi/2$ for small $\theta$.

The sum rules derived from this model will be our sum rule 2 (SR 2).

\subsection{A model with a modular $\boldsymbol{S_4}$ symmetry}
\label{sec:S4}

In \cite{Novichkov:2018ovf}  a model with a modular $S_4$
symmetry  is presented, where the authors 
mostly discussed the general case without residual symmetries.
But they also entertain the possibility to fix moduli to special values, namely
$\langle \tau_l  \rangle = \omega =  \exp(2 \, \pi \ci/3)$ respecting a $Z_3^{ST}$ and $\langle \tau_\nu  \rangle = \ci$
respecting $Z_2^S$ in the neutrino sector.
In these cases the charged lepton sector is diagonal. The neutrino sector has a type I seesaw
with the neutrino Yukawa matrix
\begin{equation}
 Y_\nu = g P_{23}\;,
\end{equation}
and the right-handed neutrino mass matrix can be written as
\begin{equation}
M_R = \begin{pmatrix}
 4 g_1 & \left(3+\sqrt{6}\right) g_2 \omega -\sqrt{6} g_3+g_3 & \left(1+\sqrt{6}\right)
   g_2-\left(\sqrt{6}-3\right) g_3 \omega ^2 \\
 * & 4 g_2 & 4 g_1+\sqrt{6} \omega  (g_2  \omega -g_3) \\
* & * & 4 g_3 
\end{pmatrix} \;.
\end{equation}
Since the matrix is symmetric we have
labelled obvious elements with '$*$' for the sake of brevity.
These matrices have the residual $Z_2^S$ symmetry generated by
\begin{equation}
 S = \frac{1}{3} \begin{pmatrix}
 - 1 & 2 \, \omega ^2 & 2 \, \omega \\
 2 \, \omega & 2 & -\omega ^2 \\
 2 \, \omega ^2 & -\omega & 2 
\end{pmatrix} \;,
\end{equation}
which is apart from phases the same as in sec.~\ref{sec:S4Sq}.

It is then easy to find that $M_R$ can be diagonalised by
\begin{equation}
 U_{23}^T U_S^T M_R U_S U_{23} = \diag(\hat{M}_1,\hat{M}_2,\hat{M}_3) \;,
\end{equation}
where 
\begin{equation}
U_S = \begin{pmatrix}
 \sqrt{\frac{2}{3}} \zeta^{-1} & \frac{1}{\sqrt{3}} \omega & 0 \\
 \frac{1}{\sqrt{6}} \omega^2 & \frac{1}{\sqrt{3}} \omega^2 & \frac{1}{\sqrt{2}} \\
 \frac{1}{\sqrt{6}} & \frac{1}{\sqrt{3}} & \frac{1}{\sqrt{2}} \zeta^{-1}
\end{pmatrix} 
\end{equation}
is a unitary matrix with $\zeta= \exp(\ci \pi/3) = \sqrt{\omega}$ and $U_{23}$ is the 2-3 rotation introduced in sec.~\ref{sec:S4Sq}. Up to phases $U_S$
is the TBM mixing matrix.

Following an approach similar to \cite{Novichkov:2018yse}, cf. sec.~\ref{sec:exp},
we redefine the parameters to find
\begin{align}
 U_{23}^T U_S^T M_R U_S U_{23} &= c \left[ \begin{pmatrix} 
   1 + \ci \sqrt{3} & 0 & 0 \\
   0 & - \text{e}^{\ci \phi } \sin(2\theta) & \cos(2\theta) \\
   0 & \cos(2 \theta) & \text{e}^{- \ci \phi } \sin(2\theta)
  \end{pmatrix}
  \right. \nonumber\\
  &+ a \begin{pmatrix}
    - 1 + \ci \sqrt{3} & 0 & 0\\
    0 & - 2 \text{e}^{2 \ci \phi } \sin^2 \theta & \text{e}^{\ci \phi } \sin(2\theta) \\
    0 & \text{e}^{\ci \phi } \sin(2\theta) & -2 \cos^2 \theta 
   \end{pmatrix} \nonumber\\
  & + b \left. \begin{pmatrix}
    0 & 0 & 0\\
    0 & 2 \cos^2 \theta  & \text{e}^{- \ci \phi } \sin(2\theta) \\
    0 & \text{e}^{- \ci \phi } \sin(2\theta) & 2 \text{e}^{2 \ci \phi } \sin^2 \theta
   \end{pmatrix} \right] \;.
\end{align}
This matrix can be made diagonal by choosing $\theta$ and $\phi$
appropriately and we will use that
\begin{equation}
b = - \text{e}^{\ci \phi} \cot (2 \theta) - a \, \text{e}^{2 \ci \phi } \;.
\end{equation}
The light neutrino mass matrix is given by the seesaw formula
\begin{align}
 m_\nu &= - v_u^2 Y_\nu^T M_R^{-1} Y_\nu = - v_u^2 g^2 P_{23} U_S U_{23} \diag(\tilde{M}_1,\tilde{M}_2,\tilde{M}_3)^{-1} U_{23}^T U_S^T P_{23}^T  \;.
\end{align}
The PMNS matrix is therefore given by
\begin{align}
 U_{\text{PMNS}} = P_{23} U_S^* U_{23}^* \Gamma_i \;,
\end{align}
where $\Gamma_i$ is again a diagonal phase matrix and
$U_{\text{PMNS}}^T m_\nu U_{\text{PMNS}} = \diag(m_1,m_2,m_3)$.
Again the complex masses from the light and heavy neutrinos are related to each other
\begin{align}
 - g^2 v_u^2 \tilde{m}_1^{-1} &= \tilde{M}_1 = c \left( a (-1 + \ci \sqrt{3}) + 1 + \ci \sqrt{3} \right) \;, \\
 - g^2 v_u^2 \tilde{m}_2^{-1} &= \tilde{M}_2 = c \left( - 2 \, a \, \text{e}^{2 \ci \phi } - \text{e}^{\ci \phi } \cot \theta  \right)  \;, \\
 - g^2 v_u^2 \tilde{m}_3^{-1} &= \tilde{M}_3 = c \left( - 2 \, a + \text{e}^{- \ci \phi } \tan \theta  \right)\;,
\end{align}
from which we can derive the mass sum rule coefficients
\begin{align}
f_1&=\frac{2/(\cos \theta \sin \theta)}{(-2 - 2 \ci \sqrt{3}) \text{e}^{\ci \phi} + 
 \ci (\ci + \sqrt{3})\cot\theta} 
  = \frac{1}{  \omega^2 \, \text{e}^{\ci \phi} \sin(2 \theta) + \omega \, \cos^2\theta} 
 ~,\\
f_2&= -\frac{(\ci + \sqrt{3} + 2 (-\ci + \sqrt{3}) \text{e}^{\ci \phi} \cot\theta) \tan\theta}{
 2 (-\ci +\sqrt{3}) \text{e}^{3 \ci \phi} - (\ci + \sqrt{3}) \text{e}^{2 \ci \phi} \cot\theta} 
 = -\frac{(1 + \omega) \tan \theta  +  2 \, \text{e}^{\ci \phi} }{
  2 \, \text{e}^{3 \ci \phi} - (1+\omega) \text{e}^{2 \ci \phi} \cot\theta} ~,\\
d&=-1 \;.
\end{align}
The relations for the mixing angles and the Dirac CP phase
are straightforward to derive as in sec.~\ref{sec:S4Sq}
\begin{align}
 \sin \theta_{13} &= \frac{1}{\sqrt{3}} \sin \theta \;,\\
 \tan \theta_{12} &=  \frac{1}{\sqrt{2}} \cos \theta \;,\\ 
 \tan \theta_{23} &= \left| \frac{2  \, \text{e}^{\ci \phi}  \tan \theta  + \sqrt{3/2} \left(1+\ci \sqrt{3}\right) }
   {3\sqrt{2/3} -  \left(1-\sqrt{3} \ci \right) \text{e}^{\ci \phi} \tan \theta}\right|  \;,\\ 
  \tan \delta &=  - \frac{(\cos (2 \theta)+5) \left(\sqrt{3} \sin \phi - 3 \cos\phi \right)}{(5 \cos (2 \theta)+1) \left(\sqrt{3} \cos \phi+ 3 \sin \phi \right)} \;.
\end{align}
To obtain maximal $\theta_{23}$, $\phi=5 \pi/6$. Again since
$\theta$ is small and with $\phi=5\pi/6$ the CPV phase is given
as  $\delta\approx \pm \pi/2$ for maximal $\theta_{23}$.

The sum rules derived from this model will be our sum rule 3 (SR 3).

\subsection{A model with modular $\boldsymbol{A_5}$ symmetry}

In \cite{Novichkov:2018nkm}  two $A_5$ models were studied, from which one model is already
excluded. 
The authors only discuss Weinberg operators and the charged lepton sector is diagonal.
For the remaining model they set $\langle \tau_\nu \rangle = \ci$ such that a $Z_2^S$ symmetry
is preserved. The generator $S$ can be made diagonal using golden ratio mixing \cite{Everett:2008et}
\begin{equation}
U_{\text{GR}}=
\begin{pmatrix}
	\sqrt{\frac{\phi_g}{\sqrt{5}}} & \sqrt{\frac{1}{\phi_g\sqrt{5}}} & 0\\
	-\sqrt{\frac{1}{2\phi_g\sqrt{5}}} & \sqrt{\frac{\phi_g}{2\sqrt{5}}} & \frac{1}{\sqrt{2}}\\
	-\sqrt{\frac{1}{2\phi_g\sqrt{5}}} & \sqrt{\frac{\phi_g}{2\sqrt{5}}} & -\frac{1}{\sqrt{2}}
	\end{pmatrix} \;,
	 \label{eq:U_GR}
\end{equation}
where $\phi_g = (1+ \sqrt{5})/2$
which also puts the neutrino matrix into block-diagonal form
\begin{align}
 U_{\text{GR}}^T m_{\nu} U_{\text{GR}} = c \begin{pmatrix} 
  b & 0 & 1 \\
  0 & \frac{1}{2 \sqrt{5}} \left( (1+\sqrt{5}) a - (1 - \sqrt{5}) b - 8  \right) & 0\\
  1 & 0 & -a
 \end{pmatrix} \;,
\end{align}
after a reparametrisation. We can now apply a 1-3 rotation as in eq.~\eqref{eq:TBM13}
and with $b = - \text{e}^{\ci \phi} \cot (2 \theta) - a \, \text{e}^{2 \ci \phi }$ we find the
complex neutrino masses
\begin{align}
\tilde{m}_1 &=  c \left( -\text{e}^{\ci \phi} \cot (\theta) - a \, \text{e}^{2 \ci \phi}  \right) \;,\\
\tilde{m}_2 &= \frac{c}{10} \left( \left(2 \left(\sqrt{5}-5\right)   \text{e}^{\ci \phi} \cot (2 \theta)-8 \sqrt{5}  \right) 
+ a \left(\left(\sqrt{5}-5\right) \text{e}^{2 \ci \phi}+5+\sqrt{5}\right)\right) \;,\\
\tilde{m}_3 &=  c \left( \text{e}^{- \ci \phi} \tan (\theta) -a \right) \;,
\end{align}
leading to the mass sum rule coefficients
\begin{align}
f_1 &=  \text{e}^{-2 \ci \phi} \frac{\left(1-\sqrt{5}\right) \text{e}^{2 \ci \phi} \cot \theta+\left(\sqrt{5}+1\right) \tan
   \theta -8 \, \text{e}^{\ci \phi}}{\left(1-\sqrt{5}\right) \text{e}^{2 \ci \phi} \tan \theta+\left(\sqrt{5}+1\right)  \cot
   \theta+8 \, \text{e}^{ \ci \phi}} \;, \\
f_2 &=  \frac{10}{\left(\sqrt{5}-5\right) \text{e}^{2 \ci \phi} \sin^2 \theta + 4\sqrt{5}  \, \text{e}^{\ci \phi} \sin(2 \theta)+\left(5+\sqrt{5}\right) \cos^2 \theta} \;,\\
   d&= +1 \;.
\end{align}

For the mixing sum rules in this case we find
\begin{align}
  \sin \theta_{13} &= \sqrt{ \frac{1}{10} (5 + \sqrt{5}) }  \sin \theta \;,\\
 \tan \theta_{12} &=  \frac{2}{1 + \sqrt{5} } \frac{1}{\cos \theta} \;,\\ 
   \tan \theta_{23} &=  \left| \frac{\sqrt{ \sqrt{5}\phi_g}   -   \text{e}^{-\ci \phi} \tan \theta}{\sqrt{ \sqrt{5} \phi_g}   +   \text{e}^{-\ci \phi} \tan \theta    }\right| \;, \\
\tan \delta &= \frac{4 \sqrt{5+\sqrt{5}} \sin (\phi) \left(2 \left( \sqrt{5}+2\right) \cos ^2(\theta) +1+\sqrt{5}\right)}
  {D_\delta} \;, \\
   D_\delta &= 4 \sqrt{5+\sqrt{5}} \cos (\phi) \cos (2 \theta) \left( (\sqrt{5}+2) \cos (2 \, \theta)+3+2\sqrt{5} \right) \nonumber\\
    &+\sqrt{2} \sin (2 \, \theta) \left( (5\sqrt{5}+11) \cos (2 \theta)+19+9 \sqrt{5}\right) \cos (2 \, \theta_{23}) \;.
\end{align}
Similar as for SR 1 we find that for maximal $\theta_{23}$ the phase $\phi=\pm \pi/2$
and then $\delta=\phi$ independent from the value of $\theta$.

The sum rules derived from this model will be our sum rule 4 (SR 4).

\section{More models with sum rules: Phenomenological Results}
\label{sec:results}

In this section we present an overview of our phenomenological results. Having provided
the analytical expressions for the mass and mixing sum rules in the previous
section we evaluate them  now numerically to show their predictions for upcoming experiments and compare them to each other.

\begin{table}
\begin{center}
\begin{tabular}{l c c c c c c c c}
\toprule
SR & $d$ & $|f_1|$ & $\arg(f_1)$ & $|f_2|$ & $\arg(f_2)$ &Group & Ref. \\
\midrule
1, I (NO) & 1 &1.00& -1.68 &2.71  &0.73  &$A_4$ & \cite{Novichkov:2018yse}\\
1, I (IO) &  1 & 1.00&-1.68 &2.58 &0.73 &$A_4$ & \cite{Novichkov:2018yse}\\
1, II (NO) &1 &1.00 &1.28 &2.71 &-0.75 &$A_4$ & \cite{Novichkov:2018yse}\\
1, II (IO) & 1 &1.05 &0.09 &2.58 & -0.64&$A_4$ & \cite{Novichkov:2018yse}\\\hline
2 (NO) & -1 &0.87 &0.43 &0.41 & -1.52&$S_4\times S_4$ & \cite{King:2019vhv} \\
2 (IO) & -1 &1.13 &0.56 &0.57 & -0.79&$S_4\times S_4$ &\cite{King:2019vhv} \\\hline
3 (NO) & -1 &0.87 &-1.66 &0.42 & 2.69&$S_4$ &\cite{Novichkov:2018ovf} \\
3 (IO) & -1 &1.13 &-1.54&0.60 & -2.86&$S_4$ &\cite{Novichkov:2018ovf}\\\hline
4 (NO) & 1 &0.45&-1.29 &1.94&0.43&$A_5$ &\cite{Novichkov:2018nkm} \\
4 (IO) & 1 &0.56&-2.27 &1.13&0.31&$A_5$ &\cite{Novichkov:2018nkm} \\
\bottomrule
\end{tabular}
\caption{Overview of parameters entering the mass sum rule for the models
we considered here. The values for  $|f_i|$ and $\arg(f_i)$ are calculated with
the best fit values of the underlying model parameters. However, it should be noted that for
SR~1 cases I and II the mass sum rule cannot be fulfilled for the $\theta,~\phi$
at their best fit values in IO.
}
\label{tab:Parameters}
\end{center}
\end{table}

\begin{table}
\begin{center}
\begin{tabular}{l c c c c c c }
\toprule
No. & $\theta_{\text{bf}} [^\circ]$ & $\phi_{\text{bf}}[^\circ]$ & $\theta_{3\sigma}[^\circ]$ & $\phi_{3\sigma}[^\circ]$ &  $\chi_{\text{min}}^2$ \\
\midrule
1 (NO) & 10.5& 47.58 &[10.0-11.0]  &$[152.5-293.5]\oplus[219.7-250.7]$  &8.6\\
1 (IO) &  10.6&50.8 &[10.1-11.0]  &$[9.5-143.5]$&19.7\\\hline
2 (NO) &15.0 &103.2&[14.3-15.7]  &$[55.9-117.3]\oplus[251.1-262.3]$ &3.3 \\
&&&&$\oplus[285.9-292.6]$&\\
2 (IO) &15.0&69.1 &[14.0-15.9]  &$[50.5-122.2]$&12.1 \\\hline
3 (NO) &14.9&163.2&[14.3-15.7]  &$[115.9-177.3]\oplus[311.1-322.3]$&3.3 \\
&&&&$\oplus[345.9-353.6]$&\\
3 (IO) &  15.0&129.0 &[14.4-15.7]  &$[114.8-173.8]$&12.1\\\hline
4 (NO) &10.1&221.6&[9.7-10.6]  &$[115.1-333.4]$&3.8 \\
4 (IO) &10.1&296.1&[9.7-10.6]  &$[185.7-336.1]$&16.5 \\
\bottomrule
\end{tabular}
\caption{Overview of the ranges of the model parameters for all sum rules.
 For some sum rules there are multiple disjoint $3\sigma$ regions for $\phi$.
 We also include the minimal value of the $\chi^2$ in these models. In IO the
 minimal $\chi^2$ without sum rules is 10.8 whereas it is zero in NO.
}
\label{tab:fits}
\end{center}
\end{table}

We begin with 
tab.~\ref{tab:Parameters}  providing a summary of parameters of all mass sum rules studied
in this manuscript including the symmetry groups the models are based on.
The quoted numbers for the coefficients $f_i$ are determined for the best fit parameters of the model
which we collected in tab.~\ref{tab:fits} together with their 3$\sigma$ ranges.
We observe that for the best fit values of the model parameters the coefficients
of the sum rules are $|f_i|\sim \mathcal{O}(1)$ however as we have seen in
sec.~\ref{sec:exp} in the $3\sigma$ range of $\theta$ and $\phi$ the coefficients
can actually have a considerable range.
Furthermore,
it is notable that we only found sum rules with $d=\pm 1$ different from the conventional
mass sum rules, where other values are possible as well, c.f.~\cite{Barry:2010yk, King:2013psa, Gehrlein:2017ryu, Gehrlein:2016wlc, Gehrlein:2015ena}.

\begin{figure}
    \centering
    \includegraphics[width=0.45\linewidth]{plots/PlotSR1I} \hspace{0.2cm}
    \includegraphics[width=0.45\linewidth]{plots/PlotSR1II} \\[0.2cm]
    \includegraphics[width=0.45\linewidth]{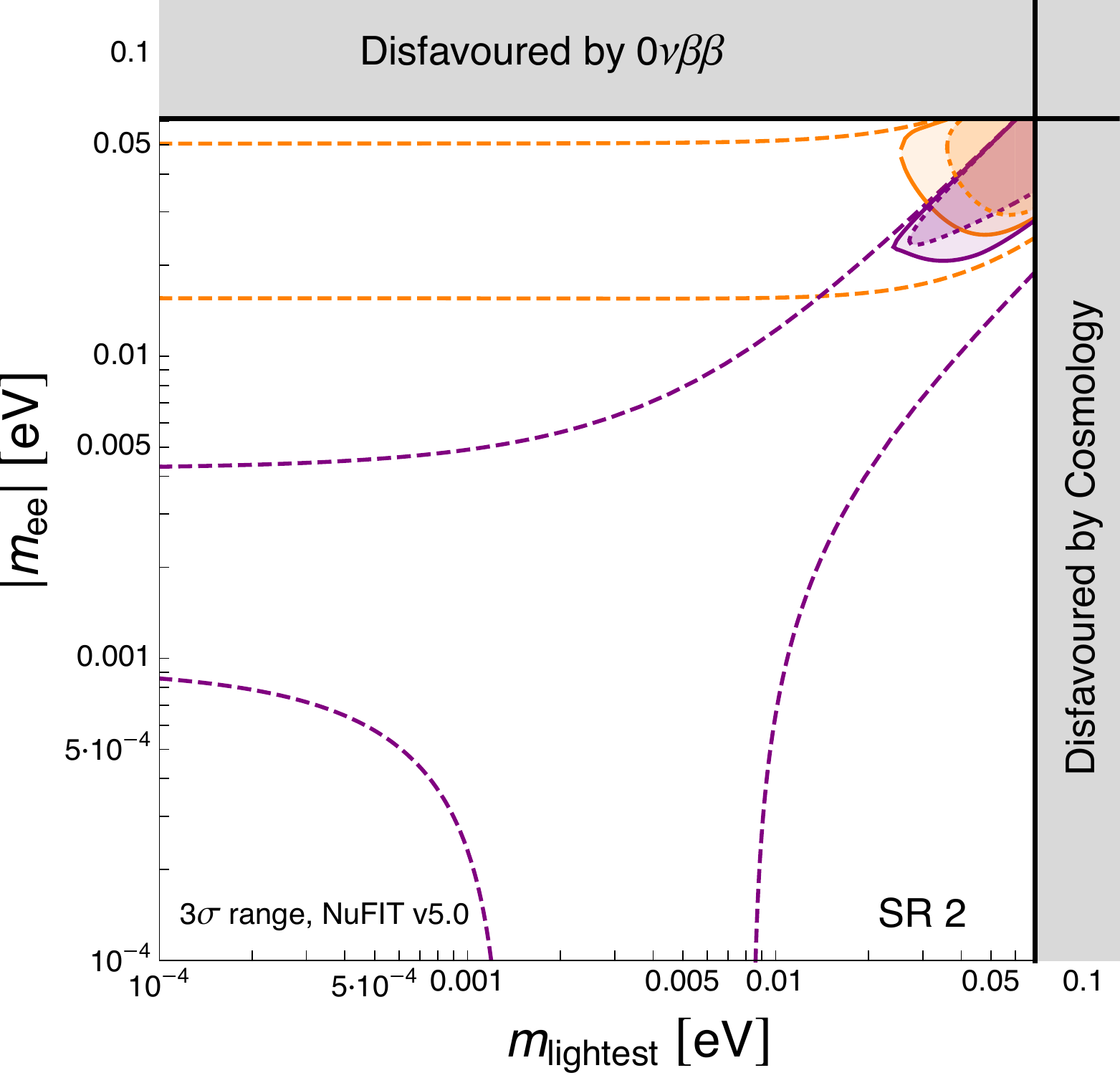} \hspace{0.2cm}
    \includegraphics[width=0.45\linewidth]{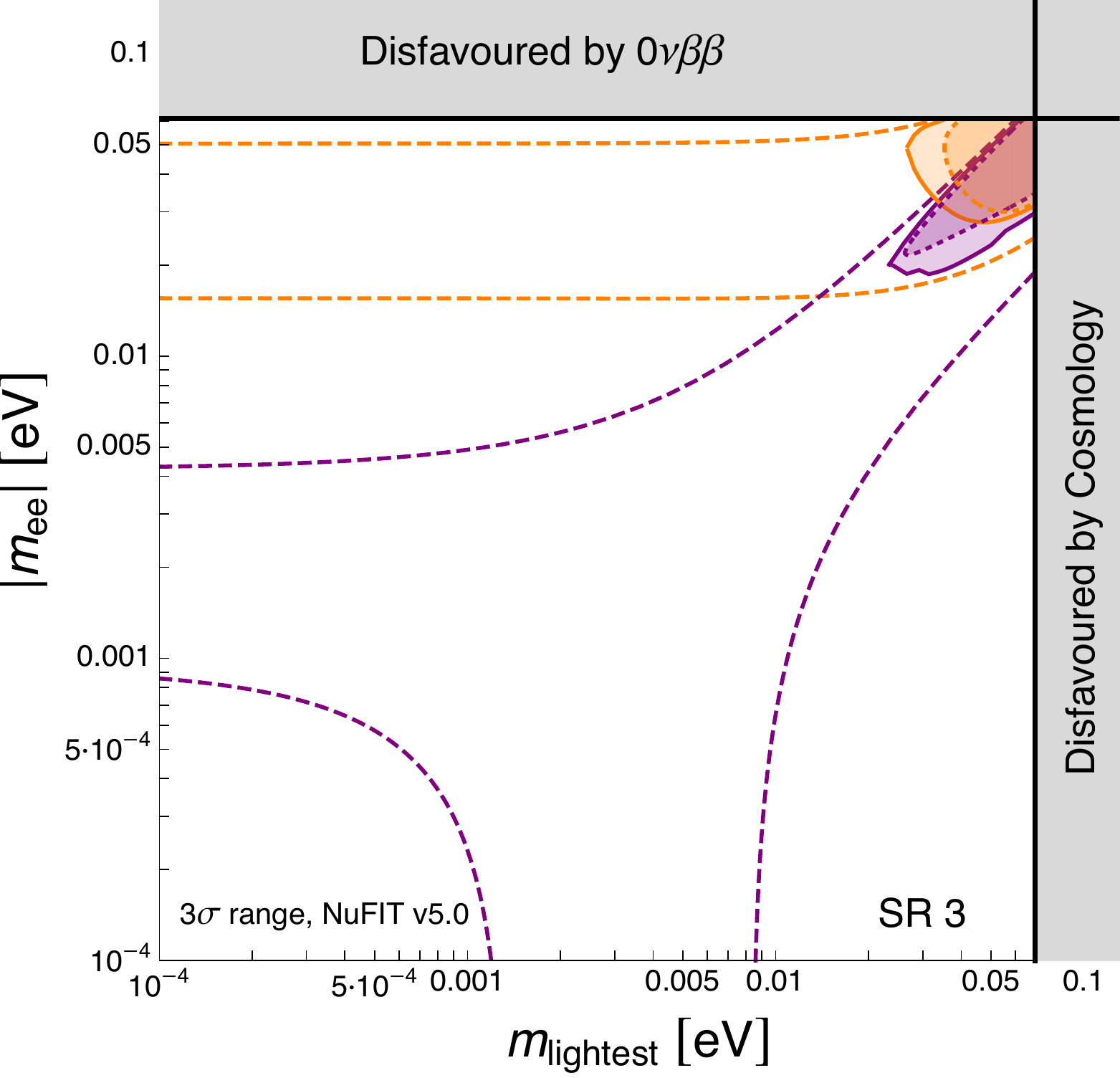} \\[0.2cm]
    \includegraphics[width=0.45\linewidth]{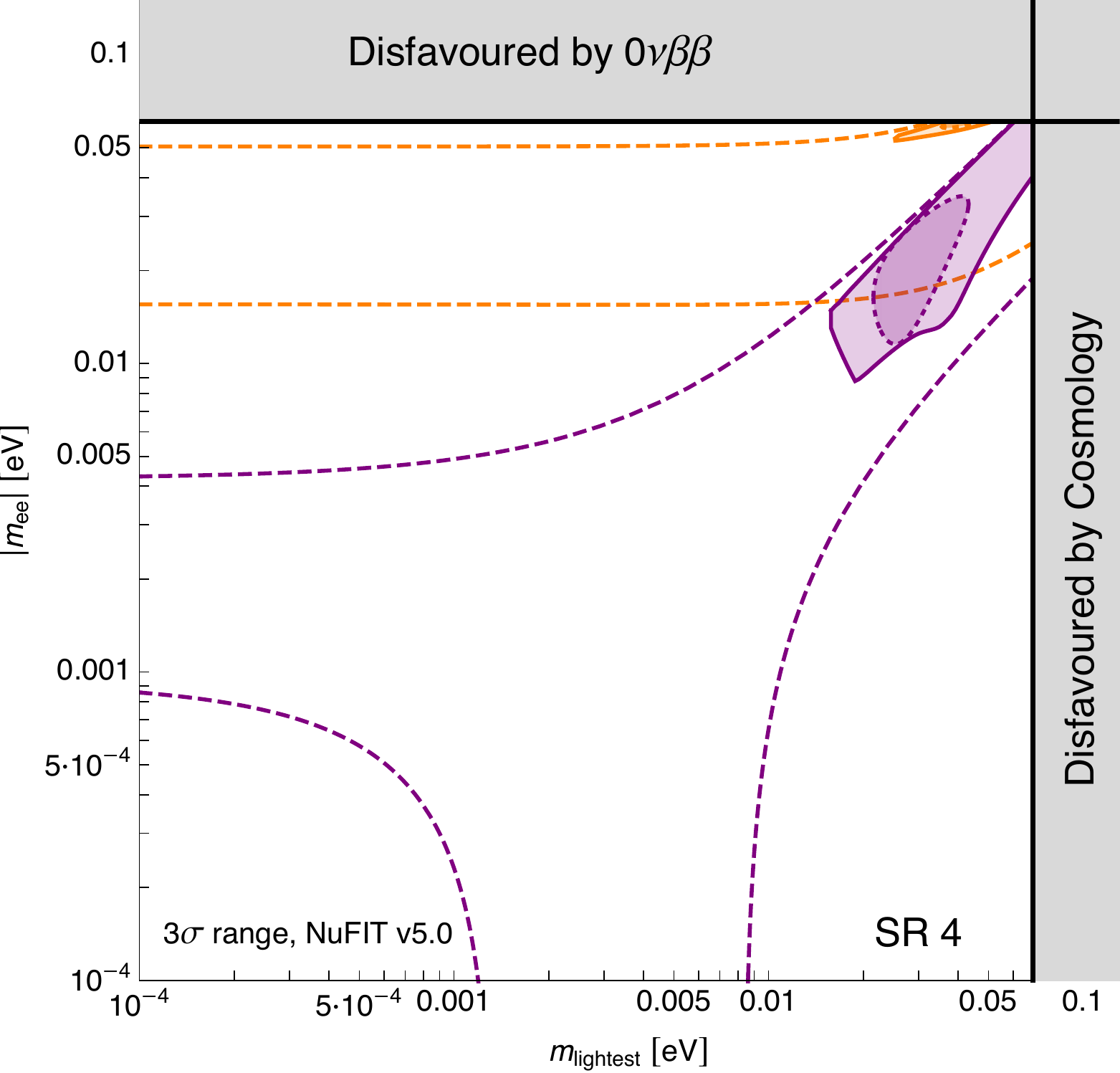}
    \caption{Allowed ranges for $|m_{ee}|$ for all sum rules.
    The purple/orange dashed region is the $3\sigma$ allowed region without
    sum rules for NO/IO. The lightly (dark) shaded purple/orange region
    is the $3\sigma$ (best fit) allowed region for NO/IO. For more details,
    see main text. 
    \label{fig:meeSRall}}
\end{figure}

\begin{table}
\begin{center}
\begin{tabular}{l l l l l c}
\toprule
No. &$m_{\text{lightest}}$& $\sum m_i$&  $|m_{ee}|$ &  $m_\beta$ & Ref. \\
\midrule
1, I (NO)	& $\geq 0$			&$\geq $57.6& $\gtrsim 0$ &$\geq $8.5& \cite{Novichkov:2018yse}\\
1, I (IO)	& $\geq 17.3$	&$\geq $124.2& $\geq 49.6$ &$\geq $52.9& \cite{Novichkov:2018yse}\\
1, II (NO)	& $\geq 0$&$\geq $57.6			& $\gtrsim 0$ &$\geq $8.5& \cite{Novichkov:2018yse}\\
1, II (IO)	& $\geq 17.3$	&$\geq $124.2& $\geq 16.7$ &$\geq $52.9& \cite{Novichkov:2018yse}\\\hline
2 (NO)	& $\geq 24.1$&$\geq $104.5	& $\geq 20.7$ &$\geq $25.5& \cite{King:2019vhv}\\
2 (IO)	& $\geq 25.4$&$\geq $138.6	& $\geq 25.2$ &$\geq $56.1& \cite{King:2019vhv}\\\hline
3 (NO)	& $\geq 23.4$	&$\geq $102.8& $\geq 18.7$ &$\geq $24.9&\cite{Novichkov:2018ovf}\\
3 (IO)	& $\geq 26.8$&$\geq $141.3	& $\geq 26.8$ &$\geq $56.7& \cite{Novichkov:2018ovf}\\\hline
4 (NO)	& $\geq 15.3$	&$\geq $84.3& $\geq 8.8$ &$\geq $17.4& \cite{Novichkov:2018nkm}\\
4 (IO)	& $\geq 24.8$&$\geq $137.5& $\geq 52.7$ &$\geq $55.7& \cite{Novichkov:2018nkm}\\
\bottomrule
\end{tabular}
\caption{Overview of lower bounds on various observables derived from the new
mass sum rules discussed in this paper. Masses are given in meV.
For more details, see main text.
}
\label{tab:Observables}
\end{center}
\end{table}

Fig.~\ref{fig:meeSRall} shows the allowed ranges for $|m_{ee}|$
over the lightest neutrino mass, $m_{\text{lightest}}$,
for all sum rules considered in this manuscript.
Additionally, we show in tab.~\ref{tab:Observables} 
the lower bounds for the lightest neutrino mass, the sum of
the neutrino masses,  $|m_{ee}|$, and $m_\beta$ in the models.
A general prediction of  mass sum rules in models with modular symmetries are
rather large values for the lightest neutrino mass $m_{\text{lightest}}>0.01$~eV,
which has also been noticed  in \cite{Feruglio:2017spp}. 
Our results do not only support this observation but they also provide a rationale for this.
Namely, the existence of mass sum rules which usually can only be fulfilled for rather large 
mass scales as for $\mathcal{O}(1)$ coefficients the neutrino masses need to be of similar order
to fulfill the sum rule. This statement however changes
 for certain parameter regions where
the coefficients can be drastically
different from $\mathcal{O}(1)$ as we have seen for SR~1 in sec.~\ref{sec:exp}.

Furthermore, SR~2 and 3 do not predict an upper limit on the mass scale in the best fit
region for $\theta$ and $\phi$ unlike SR~1 and SR~4 (see figs.~\ref{fig:meeSR1} and \ref{fig:meeSRall}).
Related to the preference for large neutrino mass scales only SR 1 cases I and II predict
values of $|m_{ee}|$ below $10^{-3}$~eV which are generically
difficult to probe experimentally.
The predicted parameter space for  SR 2, 3, 4 and parts of the parameter space for SR 1  lie within the testable
range of future
experiments for neutrinoless double beta decay which will probe $|m_{ee}|\gtrsim (10-50)$~meV
\cite{Barabash:2019suz} as well as cosmological observations of the sum of the neutrino masses
for example with CMB-stage 4 \cite{Abazajian:2016yjj} which aims to constrain the sum of the
neutrino masses below 0.5~eV, and for experiments measuring  the kinematic neutrino mass 
making our results possible benchmark points for these experiments.

\begin{figure}
    \centering
    \includegraphics[width=0.45\linewidth]{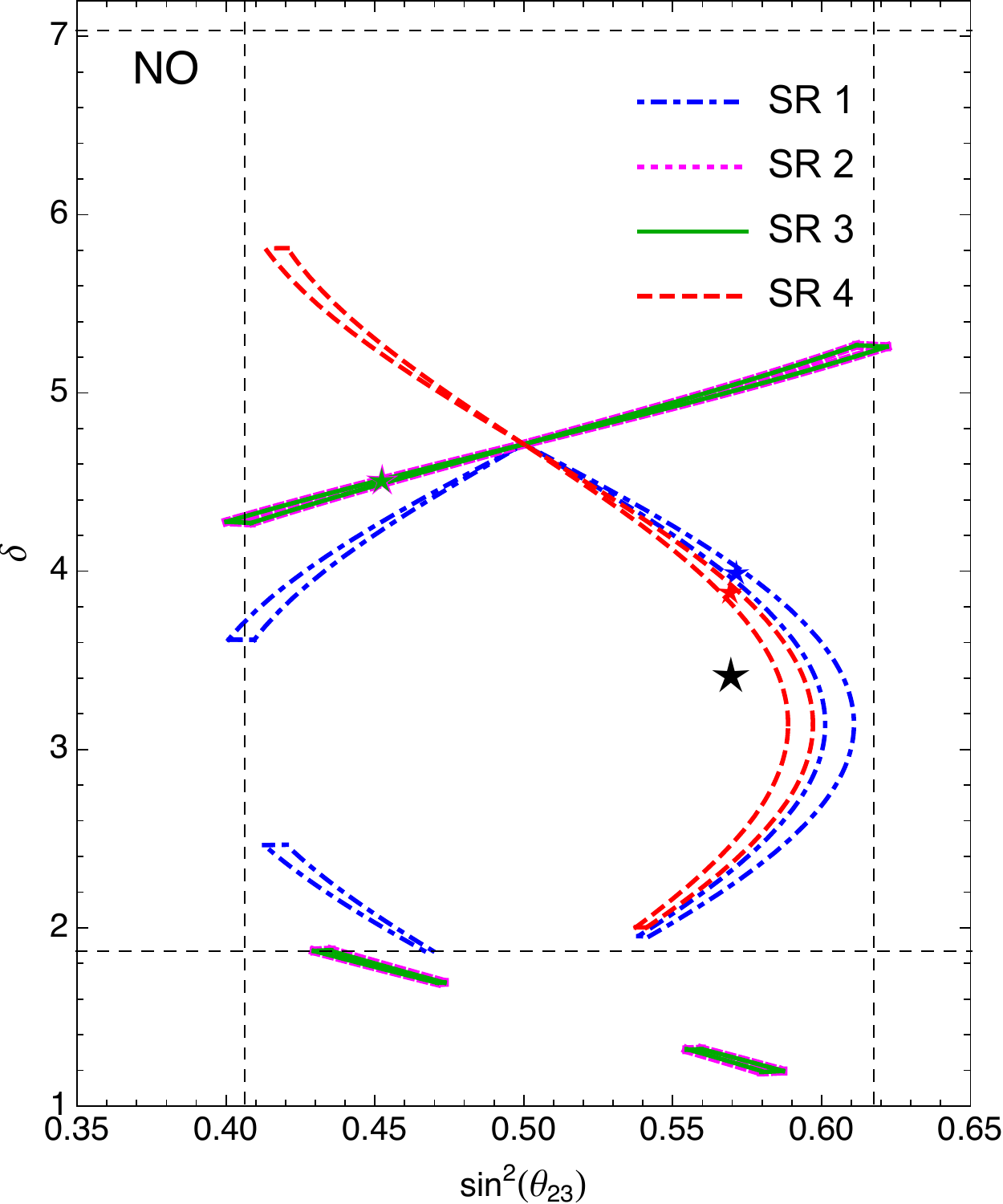} \hspace{0.2cm}
     \includegraphics[width=0.45\linewidth]{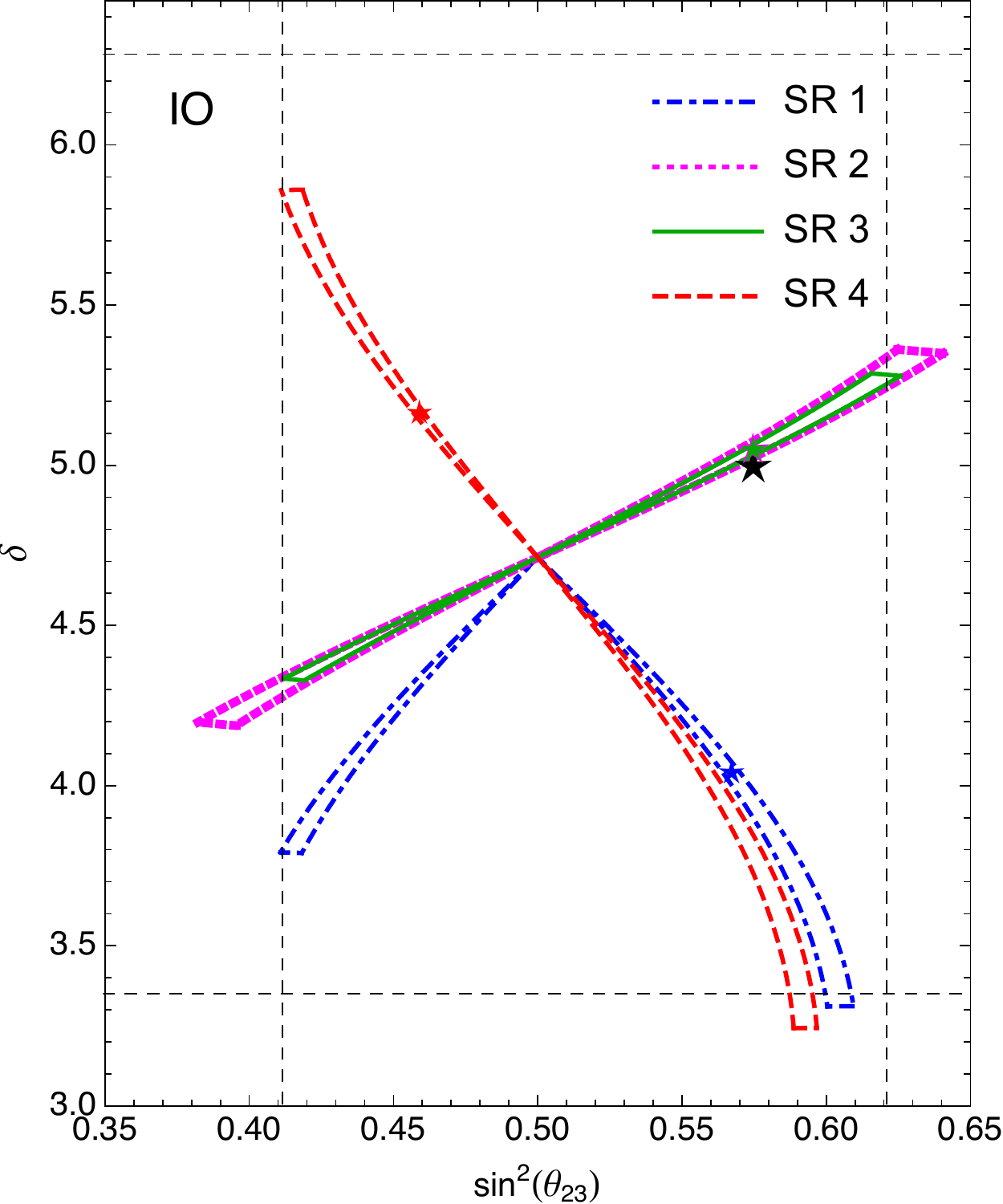} \\[0.5cm]
      \includegraphics[width=0.45\linewidth]{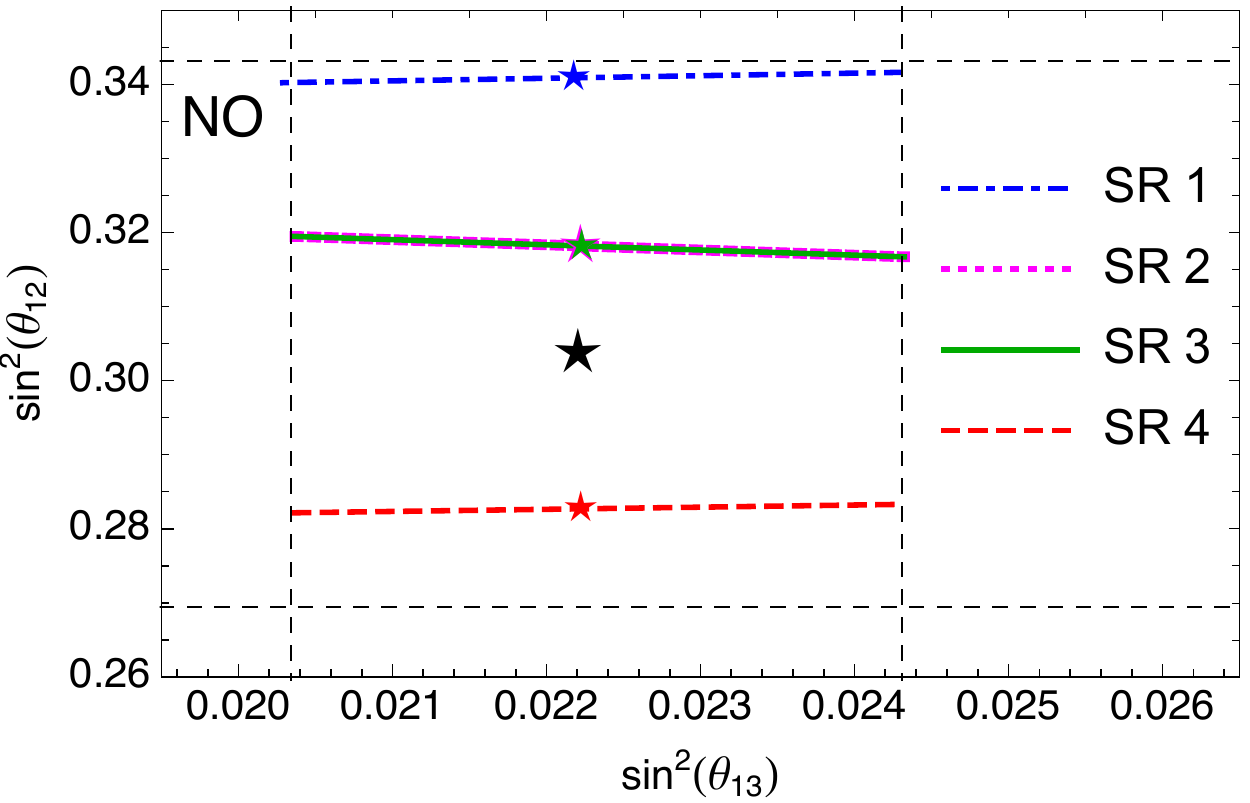} \hspace{0.2cm}
    \includegraphics[width=0.45\linewidth]{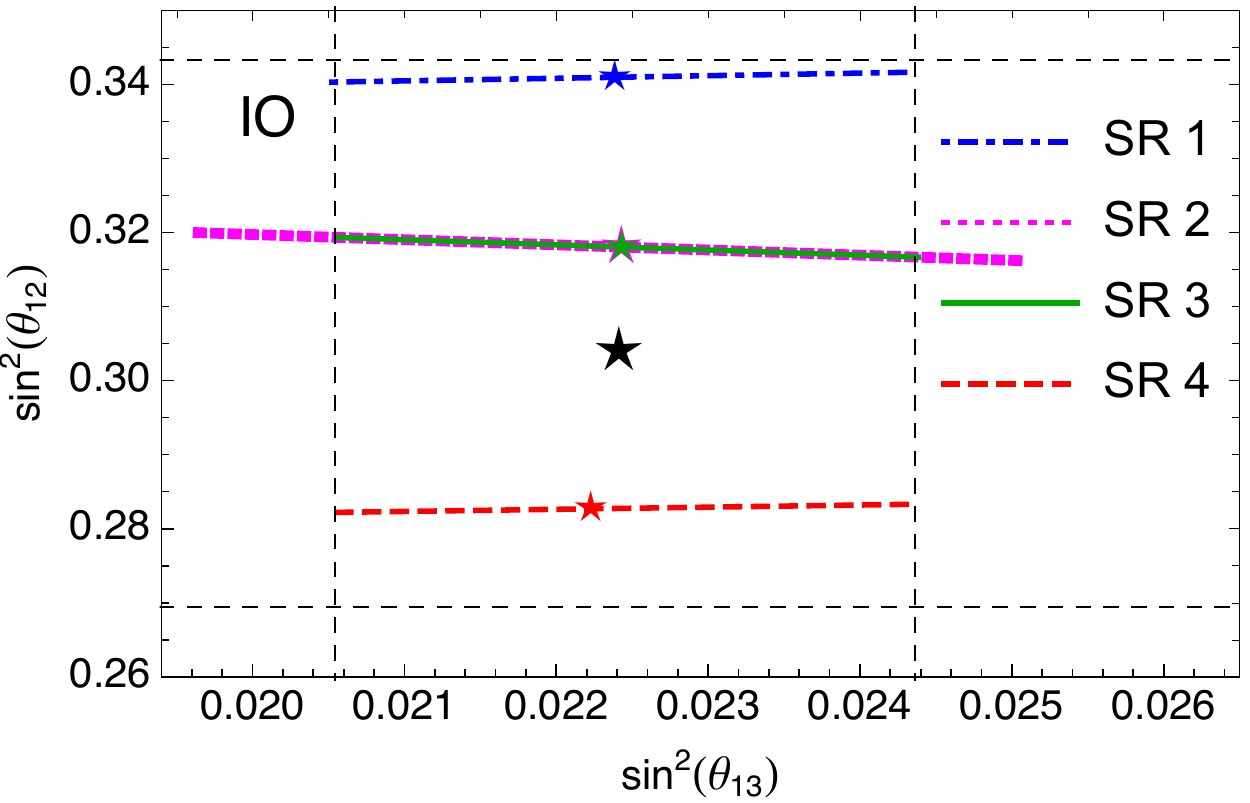}
    \caption{Allowed ranges for the mixing parameters for all sum rules  using $\theta,~\phi$ in their $3\sigma$ ranges.
    The black dashed lines show the $3\sigma$ allowed region without
    sum rules for NO/IO.  The coloured (black) stars shows the best fits with (without) sum rules.
    \label{fig:mixingSR}}
\end{figure}

Turning now to the mixing sum rules in the models considered. In fig.~\ref{fig:mixingSR} 
we show the correlations $\sin^2\theta_{12}$ vs.\ $\sin^2\theta_{13}$ and
$\sin^2\theta_{23}$ vs.\ $\delta$ respectively for both orderings. It should be noted that SR 1, case I and case II exhibit the same mixing sum rules such 
that we will refer to both cases simply as SR 1.
For all sum rules the predicted values of $\sin^2\theta_{12}$ deviate from
the currently preferred value. In the near future JUNO \cite{An:2015jdp} will
measure $\theta_{12}$ with a very good accuracy which will thoroughly probe
these models. Due to the intimate relation between $\theta_{13}$ and $\theta_{12}$
here a precise measurement of one of these angles ultimately fixes the
prediction for the other angle.

The situation for $\delta$ and $\theta_{23}$ is slightly different. Although also
in this case we obtain strong correlations there
are larger non-trivial regions allowed in the $\sin^2\theta_{23}-\delta$ plane.
Furthermore degeneracies are present such that a measurement of one parameter does
not fix the other parameter uniquely in the considered cases. For example, for SR 1 in NO
values of $\delta$ between 3.6 and 4.6 are allowed for two different 
values of $\sin^2\theta_{23}$ in both octants due to a symmetry around
$\sin^2\theta_{23}=0.5$ in the expressions for $\delta$. 
In all cases the best fit value for $\delta$ is larger than $\pi$. And $\theta_{23}$
is preferred in the lower octant for SR 2 and 3 or in the upper octant for SR 1 and 4 in NO
whereas in IO only SR 4 prefers $\theta_{23}$ in the lower octant.
Maximal $\theta_{23}$ corresponds to $\delta$ equal or close to $3\pi/2$ in all
models which can be easily understood analytically from the formulas in sec.~\ref{sec:exp}.
Furthermore, $\delta=\pi$ is only possible in NO for SR 1 and 2 for $\theta_{23}$ in
the upper octant while in IO $\delta=\pi$ is excluded. Due to the input from the global
fit, $\delta=0,~2\pi$ is not possible in any model independent
of the mass ordering, and we find a preference for CPV in all models
under consideration.

For easier comparison we also quote here for the Jarlskog invariant \cite{Jarlskog:1985ht}
\begin{align}
J_{\text{CP}}=\sin\theta_{13}\sin\theta_{12}\sin\theta_{23}\cos^2\theta_{13}\cos\theta_{12}\cos\theta_{23}\sin\delta
\end{align}
 the $3\sigma$ ranges (in units of $10^{-2}$) for the considered models
 \begin{align}
J_{\text{CP}}^{\text{NO}}\in [-3.60,3.40] \,, &\quad J_{\text{CP}}^{\text{IO}}\in [-3.61,-0.55]~\text{for SR 1}~,\\
J_{\text{CP}}^{\text{NO}}\in [-3.54,3.51] \,, &\quad J_{\text{CP}}^{\text{IO}}\in [-3.59,-2.47]~\text{for SR 2}~,\\
J_{\text{CP}}^{\text{NO}}\in [-3.54,3.50] \,, &\quad J_{\text{CP}}^{\text{IO}}\in [-3.54,-2.67]~\text{for SR 3}~,\\
J_{\text{CP}}^{\text{NO}}\in [-3.43,3.10] \,, &\quad J_{\text{CP}}^{\text{IO}}\in [-3.43,-0.32]~\text{for SR 4}~.
 \end{align}

\begin{figure}
  \centering
   \includegraphics[width=0.45\linewidth]{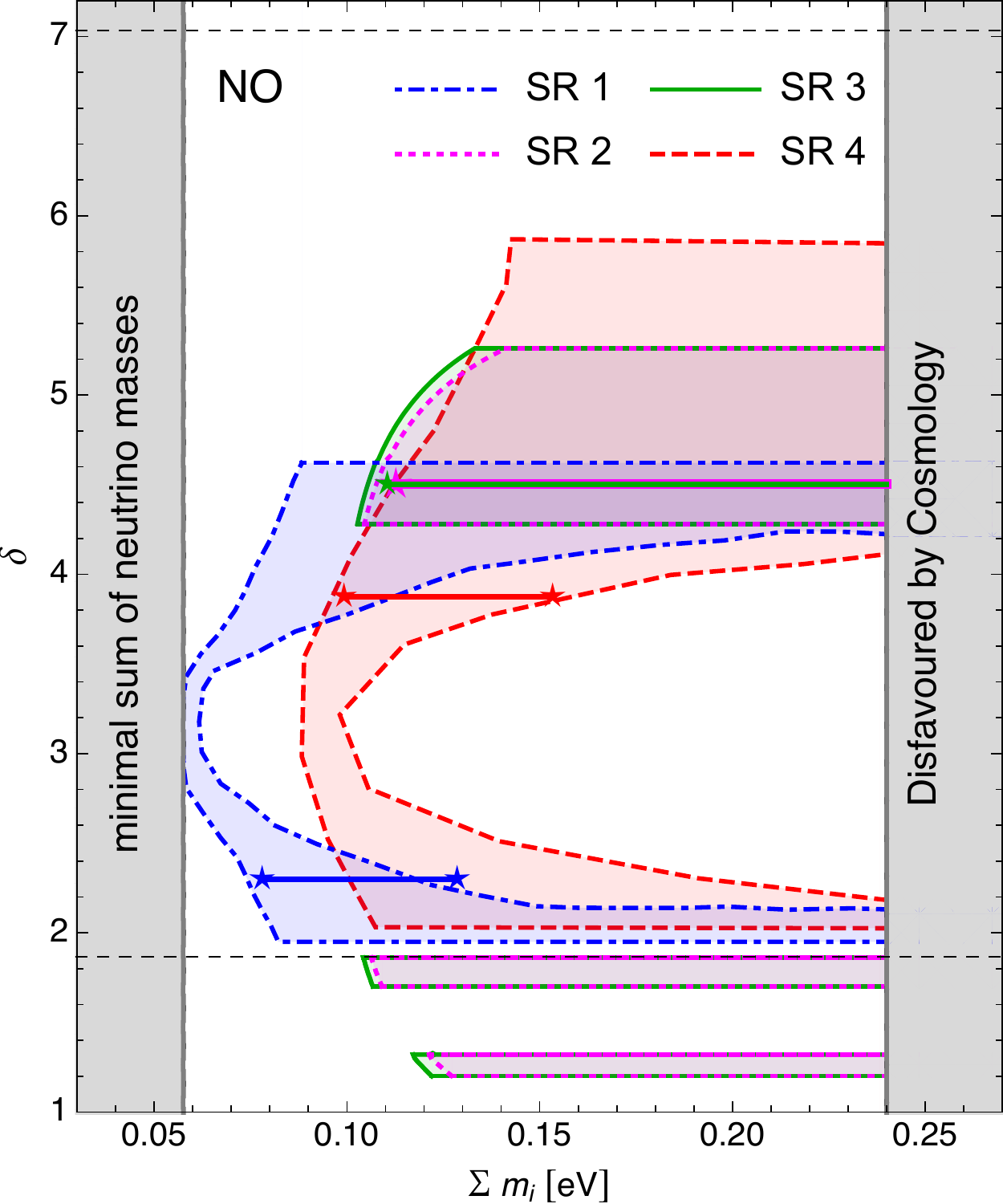} \hspace{0.2cm} 
   \includegraphics[width=0.45\linewidth]{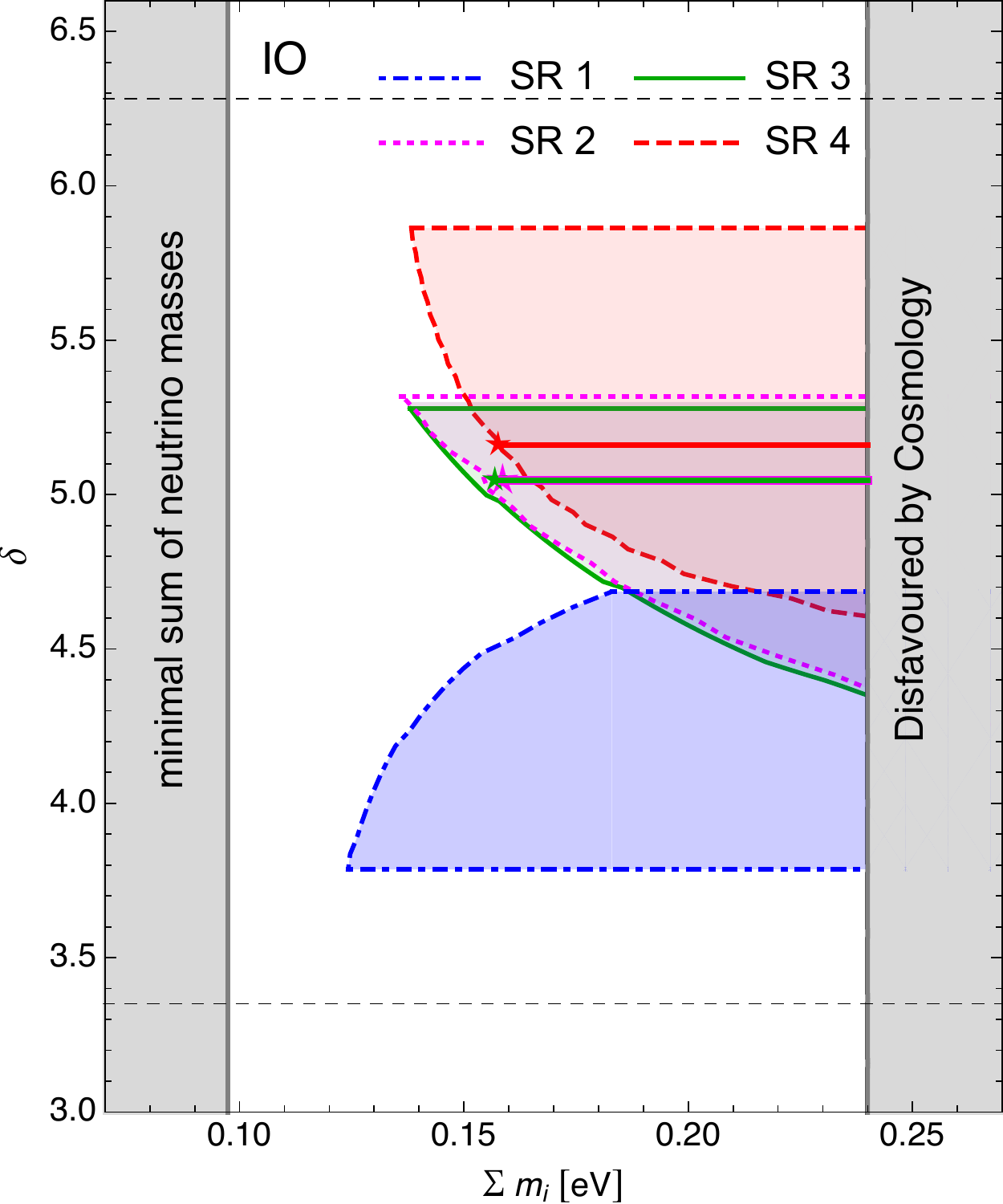} \\[0.5cm]
   \includegraphics[width=0.45\linewidth]{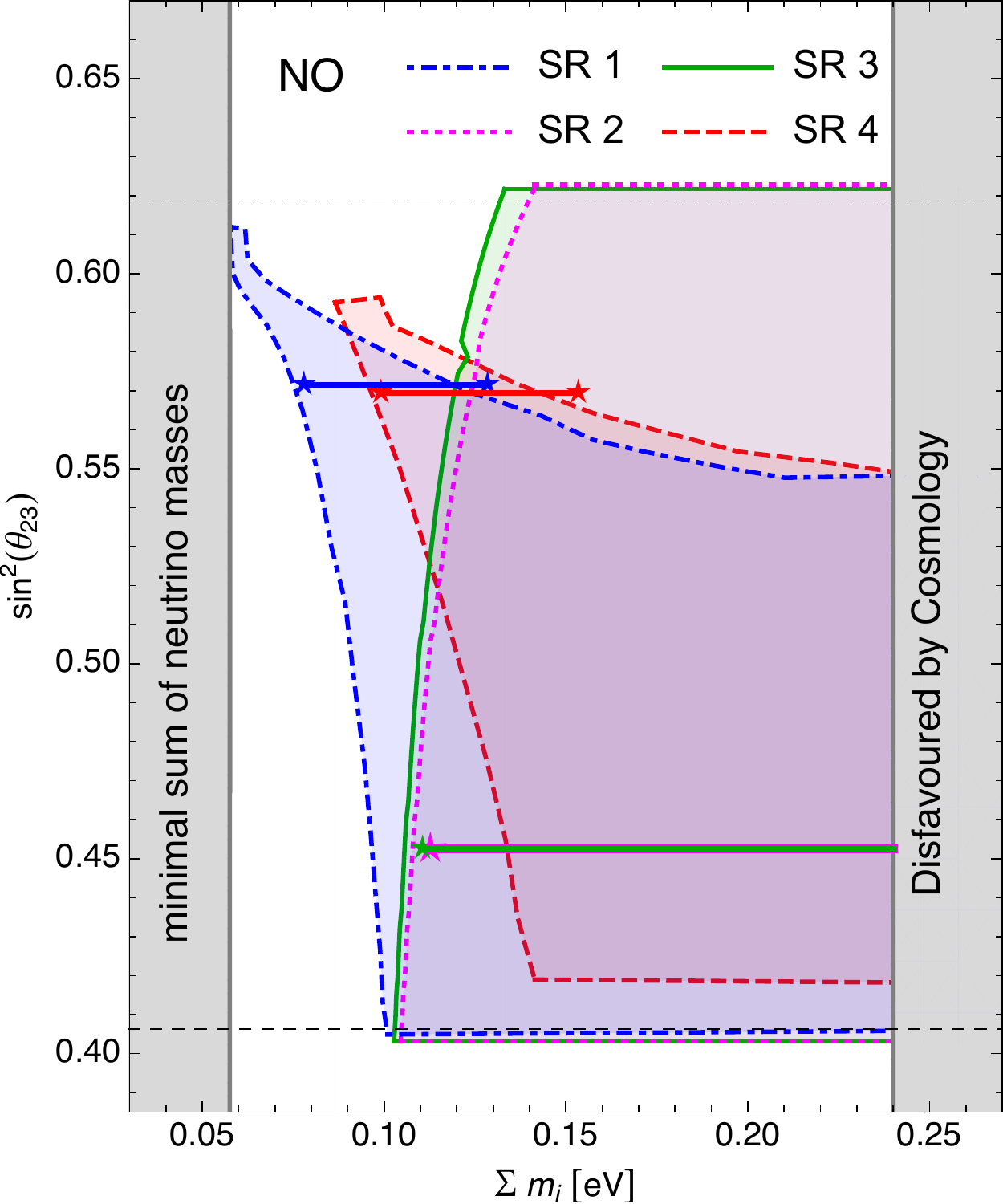} \hspace{0.2cm} 
   \includegraphics[width=0.45\linewidth]{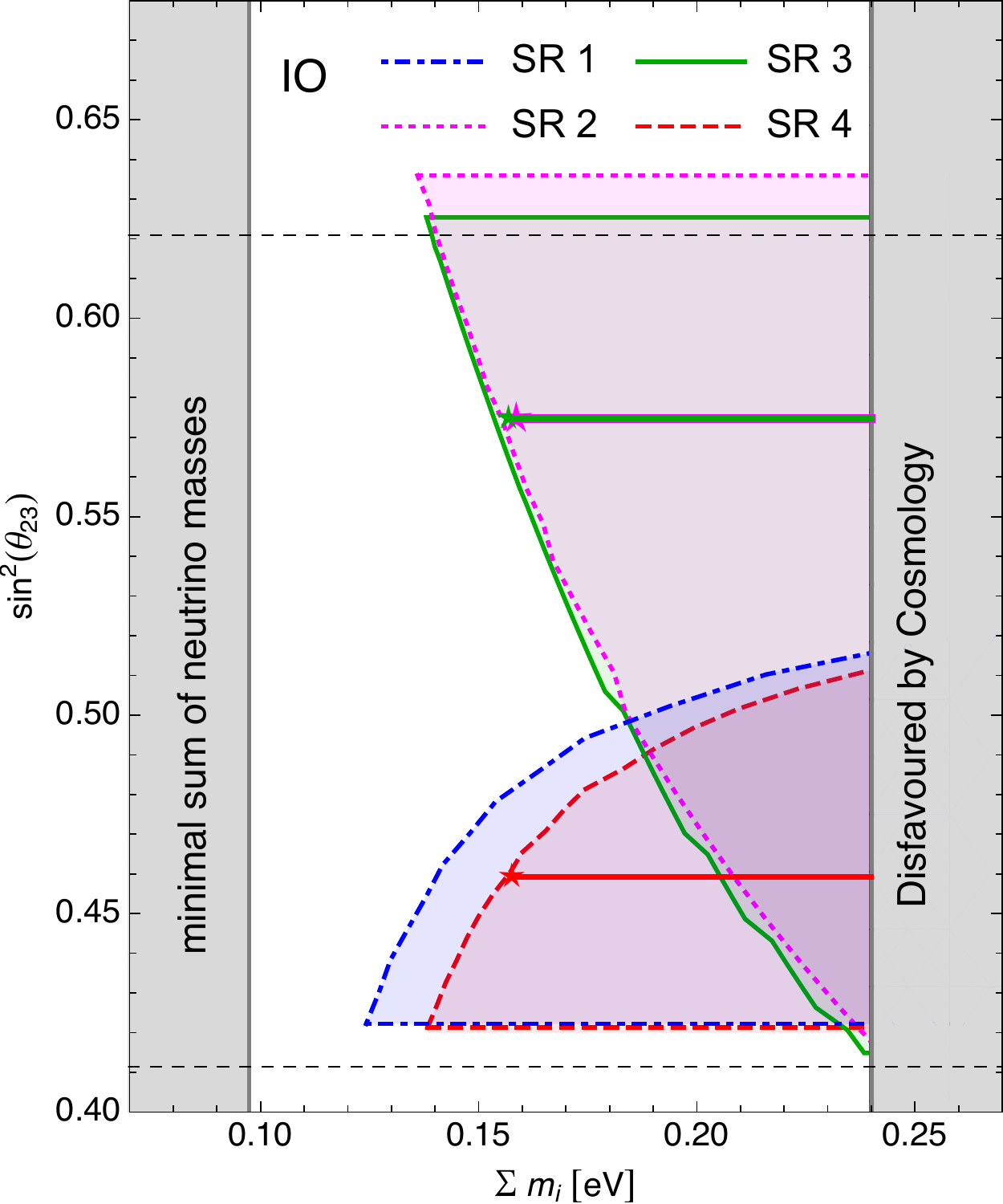} 
   \caption{
      Allowed ranges for the mixing parameters depending on the sum of the
      neutrino masses  for all SR in NO (left panels) and IO (right panels)
      for $\theta,~\phi$ and the mass splittings in their $3\sigma$ ranges.
      The stars and thick lines show the best fit regions in these models.
      Notice that for SR 1, IO the mass sum rule cannot be fulfilled if
      $\theta,~\phi$ are at best fit. The black dashed lines show the
      $3\sigma$ allowed region without sum rules for NO/IO. The gray
      exclusion regions show the minimal value of the sum of the neutrino
      masses from oscillation experiments and the maximal value of the sum
      of the neutrino masses from cosmology.
      \label{fig:mixingmassall}}
\end{figure}

We now come to the non-trivial interplay between mass and mixing
sum rules in the considered models.
In fig.~\ref{fig:mixingmassall}
we show the dependence of $\sin^2\theta_{23}$ or $\delta$ on the sum of the neutrino masses
in the $3\sigma$ ranges of the model parameters and the mass splittings.
Interestingly, we observe similar dependencies and correlations in all cases.

SR 1 case I and II exhibit the same dependencies between $\sin^2\theta_{23}$ or $\delta$
and $\sum m_i$ such that we refer to both cases as SR 1 in the following.
This is easy to understand since in  both cases the  mixing sum rules are the same
and the condition that the mass sum rule needs to be fulfilled, cf.~eq.~\eqref{eq:MassScaled1},
only depends on the absolute values of $f_1$ and $f_2$ which are identical for both cases
since they only differ in a phase. The situation is however different when deriving the predictions
for $|m_{ee}|$ which explicitly depends on the phases of $f_1$ and $f_2$, see eq.~\eqref{eq:meed1}.

For all sum rules 
a smaller mass scale leads to $\delta$ close to $\pi$ in NO such that the smallest
mass in the model can only be achieved if $\delta\approx\pi$, whereas larger neutrino masses
are only allowed for $\delta$ further away from $\pi$. However, we see
the opposite effect in IO where a smaller mass scale allows for $\delta$
closer to $2\pi$ for SR~2, 3 and 4.
Due to the
interplay of the sum rules we find a clearly preferred region
for the mass scale in the case of NO in particular for SR 1 and 4.
This also explains why in \cite{Novichkov:2020eep} the authors find
a preference for a mass scale. 
Nevertheless, in all four cases the cosmological upper bound
on the mass scale can be saturated in the 3$\sigma$ range.

Also for the relation between $\sin^2\theta_{23}$ and the sum of neutrino masses
we note different effects depending on the mass ordering.  For NO SR 2 and 3 present
only a comparatively mild dependence of the allowed range of $\theta_{23}$ on the mass scale.
For SR 1 and 4 on the other hand the dependence is stronger and we predict an upper and lower
bound on the mass scale for $\sin^2\theta_{23}>0.55$.
For IO all sum rules present strong dependency of $\sin^2\theta_{23}$ on the neutrino mass scale.
For SR 2 and 3 smaller mass scales allow for larger values of $\theta_{23}$ such that $\theta_{23}$
can only lie in the lower octant if $\sum m_i\gtrsim 0.18$ eV whereas for SR 1 and 4 the majority
of the parameter space predicts $\theta_{23}$ in the lower octant which also predicts a
smaller mass scale. In fact, $\sin^2\theta_{23}>0.5$ can only be achieved in these SR for
$\sum m_i\gtrsim 0.2$~eV (SR 4) or $\sum m_i \gtrsim 0.18 $~eV (SR 1).

It should be noted that the predictions of these sum rules have been derived assuming only small
renormalization group running effects. This is for example satisfied in the SM or in
its supersymmetric extensions for small neutrino mass scales
and small or moderate values of $\tan \beta$. However, as it has
been shown, for instance, in \cite{Gehrlein:2016fms} running effects can be large for mixing sum rules
while conventional mass sum rules are in general largely unaffected by
corrections~\cite{Gehrlein:2015ena, Gehrlein:2016wlc}.
Since the sum rules in models based on modular symmetries relate mass and mixing sum
rules the previous statement needs to be reevaluated and we expect that running effects could
be sizeable for mass sum rules in these cases as well.

\section{Summary and Conclusions}
\label{sec:sc}

In this manuscript we have studied a new class of leptonic sum rules derived
from flavour models based on modular symmetries. Specifically, we limited
this study to models  where a residual symmetry in the lepton sector is preserved.
After deriving the analytical
expressions for the sum rules we evaluated them numerically to show their predictions
for upcoming neutrino experiments.

Due to the parameter reduction in these models we find relations between
various observables. On the one hand there are mixing sum rules, which relate the mixing angles
and the Dirac CPV phase and on the other hand there are mass sum rules,
which connect the neutrino masses and Majorana phases. Similar relations have been already observed
before in flavour models with conventional discrete symmetries.
What is different here is that the coefficients of the mass sum rules are not constants anymore.
They depend on the mixing parameters, which we parametrized in terms of only two parameters,
an angle $\theta$ and a phase $\phi$. This feature leads to novel predictions, for example, 
we can get a preferred neutrino mass scale instead of just having a lower (and upper) bound.

This non-trivial interplay between mass and mixing sum rules leads to an interesting, 
distinct phenomenology compared to previous conventional models. Certain
aspects have been observed before, like the close relation between $\theta_{12}$
and $\theta_{13}$ or the appearance of a lower bound of the neutrino mass scale.
Other aspects are new. In particular, we would like to highlight here the correlation between the
Dirac CPV phase and the atmospheric mixing angle to the allowed range for the
neutrino mass scale. 
Some of our results have already been uncovered in purely numerical studies of these
models but our analytical expressions provide better insights on the origin of these results.
Hence, our work advances the study 
of  models with modular symmetries from a theoretical point of view.

This paper  demonstrates again the power of sum rules. It allowed us to write the
predictions of four very different models in a unified framework such that we can
directly compare them in terms of phenomenological predictions. That means
sum rules are an ideal tool to provide benchmark scenarios for experimental studies.
Due to their broad impact on various experiments ranging from
oscillation experiments to neutrino experiments weighing the neutrino mass scale
leptonic sum rules provide a variety of testable signatures.
Our results here can
also be immediately compared to previous systematic studies on
mass sum rules, for instance, \cite{Barry:2010yk, Dorame:2011eb, King:2013psa, Agostini:2015dna, Gehrlein:2015ena, Gehrlein:2016wlc}.
This work here shows how we can generalise that language to a completely
new class of models with modular symmetries, which have recently been investigated
in the literature.

Due to the insights provided in this manuscript, the improved comparability of results
and the ease of application in experimental
studies we hope that it will become a standard way to present predictions from flavour
models in the future.

\section*{Acknowledgment}
\label{sec:Acknowledgment}

We would like to thank Serguey Petcov for some very useful comments on this
manuscript and insights into models with modular symmetry and Arsenii Titov for some help understanding their modular $A_5$ model.
JG is supported by the US Department of Energy under Grant Contract DE-SC0012704.
MS is supported by the Ministry of Science and Technology (MOST) of Taiwan 
under grant number MOST 107-2112-M-007-031-MY3.

\end{document}